\documentclass[12pt]{article}
\usepackage{cite}
\usepackage{enumerate}
\usepackage{ifpdf}
\usepackage{ae}
\usepackage[T1]{fontenc}
\usepackage[ansinew]{inputenc}
\usepackage{amsmath}
\usepackage{relsize}
\usepackage{amssymb}
\usepackage{tikz}
\usepackage{graphicx}
\usepackage{color}
\definecolor{darkblue}{cmyk}{0.9,0.9,0,0}
\definecolor{darkgreen}{rgb}{0,0.55,0}
\usepackage[colorlinks=true,linkcolor=darkblue,citecolor=darkblue,urlcolor=darkblue]{hyperref}
\usepackage{epsfig}
\usepackage{epstopdf}
\usepackage{graphicx}

\makeatletter
\long\def\@makecaption#1#2{
  \vskip\abovecaptionskip
  \sbox\@tempboxa{{\captionfonts #1: #2}}
  \ifdim \wd\@tempboxa >\hsize
    {\captionfonts #1: #2\par}
  \else
    \hbox to\hsize{\hfil\box\@tempboxa\hfil}
  \fi
  \vskip\belowcaptionskip}
\makeatother

\newcommand{\beq}{\begin{equation}}
\newcommand{\eeq}{\end{equation}}
\newcommand{\beqy} {\begin{eqnarray}}
\newcommand{\eeqy} {\end{eqnarray}}
\newcommand{\bsmat}{\begin{smallmatrix}}
\newcommand{\esmat}{\end{smallmatrix}}
\newcommand{\bmat}{\begin{matrix}}
\newcommand{\emat}{\end{matrix}}

\def\({\left(}
\def\){\right)}
\def\[{\left[}
\def\]{\right]}

\def\<{\langle}
\def\>{\rangle}

\usepackage{varioref}
\usepackage{makeidx}
\makeindex

\usepackage[english]{babel}
\usepackage{caption}
\usepackage{subfig}

\usepackage{tabularx}

\newtheorem{theorem}{Theorem}

\begin{document}

\thispagestyle{empty}

\renewcommand{\thefootnote}{\fnsymbol{footnote}}
\setcounter{page}{1}
\setcounter{footnote}{0}
\setcounter{figure}{0}
\begin{titlepage}

\begin{center}

\vskip 2.3 cm 

\vskip 5mm

{\Large \bf
Rademacher Expansions and the Spectrum of 2d CFT
}
\vskip 0.5cm

\vskip 15mm

\centerline{Luis F. Alday, Jin-Beom Bae}
\bigskip
\centerline{\it Mathematical Institute, University of Oxford,} 
\centerline{\it Woodstock Road, Oxford, OX2 6GG, UK}

\end{center}

\vskip 2 cm

\begin{abstract}
\noindent  A classical result from analytic number theory by Rademacher gives an exact formula for the Fourier coefficients of modular forms of non-positive weight. We apply similar techniques to study the spectrum of two-dimensional unitary conformal field theories, with no extended chiral algebra and $c>1$. By exploiting the full modular constraints of the partition function we propose an expression for the spectral density in terms of the light spectrum of the theory. The expression is given in terms of a Rademacher expansion, which converges for spin $j \neq 0$. For a finite number of light operators the expression agrees with a variant of the Poincare construction developed by Maloney, Witten and Keller. With this framework we study the presence of negative density of states  in the partition function dual to pure gravity, and propose a scenario to cure this negativity. 
\end{abstract}

\end{titlepage}


\setcounter{page}{1}
\renewcommand{\thefootnote}{\arabic{footnote}}
\setcounter{footnote}{0}

 \def\nref#1{{(\ref{#1})}}
 
\section{Introduction}

The spectrum of unitary two-dimensional conformal field theories (2d CFT) can be conveniently packed in their partition function on the torus $Z(q,\bar q)$. Modular transformations on the torus $PSL(2,\mathbb{Z}): q \to q'$ leave the partition function invariant
\begin{equation}
Z(q,\bar q) = Z(q',\bar q'),
\end{equation}
and a natural question is which constraints does this impose on the spectrum. In the simpler holomorphic case this question has a long history. Hardy and Ramanujan \cite{HardyRamanujan1} considered the generating function for partitions of $n$ 
\begin{equation}
Z(q) =q^{-\frac{1}{24}} \sum_{n=0} p(n) q^n = q^{-\frac{1}{24}}\left( 1+q+2q^2+\cdots \right)
\end{equation}
and showed that the asymptotic behaviour for $p(n)$ follows from the properties under the modular transformation $q = e^{-\beta} \to q' = e^{-\frac{4\pi^2}{\beta}}$ together with the presence of the $1 \times q^{-\frac{1}{24}}$ in the small $q$ expansion. Two decades later Rademacher revisited their argument and gave an exact expression for the Fourier coefficients $p(n)$ in terms of a series known as Rademacher expansion \cite{Rademacher1}. The method exploits the full modular invariance/covariance and consist in applying Cauchy theorem in the presence of a dense tower of essential singularities, one per each $PSL(2,\mathbb{Z})$ element. Each term in the Rademacher expansion corresponds to one of such singularities, modulo periodicity.\footnote{Hardy and Ramanujan had given a similar looking expression but crucially, their results were only asymptotic and their sum over $PSL(2,\mathbb{Z})$ was not convergent.} More generally, Rademacher proved that the Fourier coefficients of any holomorphic modular form of non-positive weight of the form\footnote{Since $Z(q)$ is not as $q \to 0$ it is more precise to denote this a {\it weakly} holomorphic modular form, but from now on we will suppress the term weakly.}
\begin{equation}
Z(q) = \sum_{n=0} p(n) q^{n - \hat c} =  \sum_{n < \hat c} p(n) q^{n - \hat c}+ \sum_{n \geq \hat c} p(n) q^{n - \hat c}
\end{equation}
are again given by a Rademacher expansion and follow from the polar terms, with $n < \hat c$, together with the full modular covariance of $Z(q)$. The punchline of the Rademacher construction is that modular forms of non-positive weight are fully fixed by their polar part.   An alternative way to construct a modular form with a given  polar part is to construct its Poincare series, namely a sum of the polar terms together with all their $PSL(2,\mathbb{Z})$ images. For holomorphic modular forms of non-positive weight these two constructions agree. 

These developments were mimicked by physicists in studying the spectrum of unitary 2d CFT and related questions. A similar argument to the one by Hardy and Ramanujan led Cardy to his celebrated formula for the asymptotic growth of states  \cite{Cardy:1986ie}, and also allows to compute microscopically the entropy of black holes with $AdS_3$ near horizon geometry, in terms of states the boundary 2d CFT \cite{Strominger:1997eq}.  Rademacher expansions were first studied in the context of 2d CFT in \cite{Dijkgraaf:2000fq}, where they were used to give an exact expression for the Fourier coefficients of elliptic genera on CY manifolds. 

In this paper we address the following question. Given the partition function of a unitary 2d CFT on the torus 
\begin{equation}
Z(q,\bar q) = \sum_{h,\bar h} q^{h-c/24} \bar q^{\bar h-c/24},
\end{equation}
where $c$ is the central charge of the theory, what are the constraints imposed by the full modular invariance. We will study this question in the context of a theory with Virasoro but not extended chiral symmetry and central charge $c>1$. In this case $\hat c = \frac{c-1}{24}$ is the natural combination appearing in the partition function. A natural question is whether the spectrum in the {\it censored region}, with either $h-\hat c<0$ or $\bar h-\hat c<0$ and which corresponds to the polar terms in the holomorphic case, determine the rest of the spectrum. The answer in the non-holomorphic case is negative. In the holomorphic case the space of modular forms of a given weight is well understood. In particular one can prove that a bounded modular form of weight zero is necessarily a constant, so that basically we can only add a constant to the Rademacher expansion. In the non-holomorphic case this is not the case, and the space of modular forms with a given weight is much less understood, see for instance \cite{Brown:2017qwo}. 

The next question we can ask is how to construct a partition function consistent with modular invariance and the presence of a given spectrum in the censored region. An example of such a partition function was constructed by Maloney and Witten  \cite{Maloney:2007ud} and further analysed by Keller and Maloney \cite{Keller:2014xba}. This partition function is the (appropriately regularised) Poincare series given by a {\it seed} with quantum numbers $(E,J)$ plus all its $PSL(2,\mathbb{Z})$ images. When the seed is in the censored region, the rest of the spectrum is in the uncensored region. The simplest example corresponds to pure gravity in $AdS_3$: in this case the only operators in the censored region are the vacuum plus all its Virasoro descendants. The resulting density of states, denoted by $\rho_P(e,j)$,  suffers from two problems: it possesses states with negative norms, and results in a continuous spectrum on the energy $e$, as opposed to discrete.   

In this paper we will follow a different approach and mimic the Rademacher construction for the non-holomorphic case. In this way we construct a density of states $\rho_R(e,j)$ consistent with the presence of a single operator in the censored region and full modular invariance. This density is given by a Rademacher expansion convergent for all spin $j \neq 0$. For large spin it reproduces the asymptotic behaviour previously found in the literature \cite{Benjamin:2019stq} while for finite spin it is given by a variant of the Maloney, Witten, Keller (MWK) Poincare density. It is interesting to note that, unlike in the holomorphic case, the Poincare and Rademacher construction lead to two slightly different answers. We show however, that they are physically equivalent: the difference between the two corresponds to the Poincare series for a density of `extra' operators in the uncensored region. This extra contribution gives an `oscillatory' contribution on top of the exponentially large terms. We argue that this is the sort of ambiguities present in a generic situation. As a byproduct of this comparison we are able to find analytic expressions  for $\rho_P(e,j)$ and $\rho_R(e,j)$. They are sums of terms of the form
\begin{eqnarray}
\rho_P(e,j) &=& \frac{\cosh \left(\sqrt{2} \pi(\zeta \kappa_+ - \frac{j \kappa_-}{\zeta})\right)+\cosh \left(\sqrt{2} \pi( \frac{j \kappa_+}{\zeta}- \zeta \kappa_-)\right)-2}{\sqrt{e^2-j^2}} \Theta(e-j), \\
\rho_R(e,j) &=& \frac{\sinh \left(\sqrt{2} \pi(\zeta \kappa_+ - \frac{j \kappa_-}{\zeta})\right)+\sinh \left(\sqrt{2} \pi( \frac{j \kappa_+}{\zeta}- \zeta \kappa_-)\right)}{\sqrt{e^2-j^2}} \Theta(e-j),
\end{eqnarray}
where $\kappa_{\pm} = \sqrt{J-E}\pm \sqrt{-E-J}$ encode the information about the seed and we have introduced the combination $\zeta =\sqrt{e+\sqrt{e^2-j^2}}$. Although we provide explicit results for isolated operators, our method can also be applied to situations with accumulation points in the twist (provided the accumulation is mild enough) and we treat some examples. For the case of pure gravity the Rademacher construction provides an equally good density, in the sense that it reproduces the known contributions to the partition function from classical geometries. On the other hand, it also suffers from the same problems. It has negative norm states and it leads to a continuous spectrum. While we don't have any proposals to render the spectrum discrete, we discuss scenarios to cure the negativity of the density. The simplest scenario involves adding a tower of extra operators, whose twist we compute.  

This paper is organised as follows. In section \ref{Rademacher} we discuss the Rademacher construction in the holomorphic case and show how it works in a few examples. In section \ref{Rademacher2dCFT} we apply these ideas to analyse the spectrum of unitary 2d CFT. In section \ref{PvsR} we discuss the issue of ambiguities, make a comparison between the Rademacher and Poincare constructions, and discuss a scenario to cure the negativity of states, together with other scenarios with accumulation points in the twist. We finally end up with a list of open problems. In the appendix we review the construction by MWK and give an analytic expression for the density arising from their Poincare construction. 

\section{From asymptotic to convergent expansions}
\label{Rademacher}

\subsection{A toy model}
Consider the generating function for partitions of $n$
\begin{equation}
Z_{part}(q) = \sum_{n=0} p(n) q^n = \frac{q^{\frac{1}{24}}}{\eta(\tau)},~~~\eta(\tau) = q^{\frac{1}{24}} \prod_{n=1}(1-q^n),
\end{equation}
where $\eta(\tau)$ is the Dedekind's $\eta-$function and $q=e^{2\pi i \tau}$. Our aim is to understand the asymptotic behaviour of the Fourier coefficients $p(n)$. This problem has a long history. It turns out $p(n)$ is not known in a closed form, nor does it satisfy any finite order recurrence. The leading asymptotic behaviour was first found by Hardy and Ramanujan \cite{HardyRamanujan1}. This can be done as follows. The Dedekind's $\eta-$function satisfies the following modular transformation
\begin{equation}
\eta\left(-\frac{1}{\tau} \right) = \sqrt{-i \tau} \eta(\tau),
\end{equation}
which implies that for complex $z$ with $Re(z)>0$
\begin{equation}
Z_{part}(e^{-z}) =\sum_{n=0} p(n) e^{-z n} = \sqrt{\frac{z}{2\pi}} e^{\frac{\pi^2}{6 z}} e^{-\frac{z}{24}} \left( 1 + {\cal O}(e^{-\frac{4\pi^2}{z}}) \right).
\end{equation}
The r.h.s. diverges exponentially as $z \to 0^+$. This behaviour cannot arise from a finite number of terms on the l.h.s., hence it must come from the tail with $n \gg 1$. In this regime we can approximate the sum by an integral and we must have
\begin{equation}
\int_0^\infty p(n) e^{-z n} dn \sim \sqrt{\frac{z}{2\pi}} e^{\frac{\pi^2}{6 z}} e^{-\frac{z}{24}}.
\end{equation}
Performing the inverse Laplace transform we then obtain
\begin{equation}
p(n) \sim \frac{2 \sqrt{3} e^{\frac{1}{6} \pi  \sqrt{24 n-1}} \left(\pi  \sqrt{24 n-1}-6\right)}{\pi  (24 n-1)^{3/2}}.
\end{equation}
Which indeed gives the asymptotic behaviour found by Hardy and Ramanujan. The manipulations we have done are justified by the following theorem from Tauberian theory, first proven by Wright \cite{Wright71}.
\begin{theorem}
\label{t1}
Suppose $Z(q) = \sum_{n=0}^\infty a_n q^n$ is a power series, analytic for $|q|<1$ and $q \not\in \mathbb{R}_{\leq 0}$, such that it satisfies the following two conditions. First
\begin{equation}
Z(e^{-z}) = z^\alpha e^{\frac{\kappa^2}{z}} \left( \sum_{s=0}^{N-1} \alpha_s z^s +{\cal O}(z^N) \right)+ \cdots,
\end{equation}
as $z=x+i y \to 0$ in the arc $Arg[z]< \delta$, with $\delta<\pi/2$, and $\alpha,\kappa,\alpha_s$ real coefficients. Here the dots denote contributions exponentially suppressed with respect to the leading one. Second
\begin{equation}
Z(e^{-z}) \ll Z(e^{-x}) e^{-\frac{d}{x}},
\end{equation}
for some $d>0$ and as $z=x+i y \to 0$ in the complementary arcs $\frac{\pi}{2} -\delta \leq Arg[z]< \frac{\pi}{2}$. Then, one can prove the following asymptotic expansion for the coefficients $a_n$ in the large $n$ limit
\begin{equation}
a_n =e^{2 \kappa \sqrt{n}} n^{-\frac{\alpha}{2}-\frac{3}{4}} \left( \sum_{r=0}^{N-1} p_r n^{-r/2}  + {\cal O}(n^{-N/2}) \right) + \cdots,
\end{equation}
where the coefficients $p_r$ are computable in terms of the coefficients $\alpha_s$ 
\begin{equation}
p_r=\sum_{s=0}^r \alpha_s w_{s,r-s},~~~~~w_{s,r}=\frac{2^{-2 r-1} \left(-\frac{1}{\kappa }\right)^r \kappa ^{\alpha +s+\frac{1}{2}} \Gamma \left(r+s+\alpha +\frac{3}{2}\right)}{\sqrt{\pi } \Gamma (r+1) \Gamma \left(-r+s+\alpha +\frac{3}{2}\right)}
\end{equation}
and dots represent contributions exponentially suppressed with respect to the leading one in the large $n$ limit. 
\end{theorem}
An alternative proof was presented in \cite{NgoR}. Very recently this was also discussed in \cite{BJM}, among other Tauberian theorems, where the importance of the second condition was stressed. Note that it is important that $Z(e^{-z})$ is defined for complex $z$, and the above conditions hold in the specified arcs. To make this point clear, suppose the asymptotic expansion contains the term $a_n = (-1)^n b_n$, where $b_n$ grows at least as fast as the contribution in the theorem.  How can we exclude the presence of such terms? For real and positive $x$ these terms would not contribute to the exponential behaviour of $Z(e^{-x})$ as $x \to 0$. However, they would invalidate the second condition of the theorem. The result of the theorem agrees with approximating the sum by an integral and taking the inverse Laplace transform. The conditions of the theorem specify under which circumstances this is the correct procedure. One can explicitly check that the example above, the generating function for partitions, indeed satisfies the conditions of the theorem. Note that the output of this theorem is stronger than that of the Tauberian theorem presented in \cite{Mukhametzhanov:2019pzy}, see also \cite{Pal:2019zzr}, where only the leading power law can be determined. The crucial difference of course is that we are considering an evenly spaced series, so that the input is stronger too. 
 
The theorem above can be generalised to problems involving alternating series with a slight modification. Suppose the following power series satisfy the asymptotics
\begin{eqnarray}
&&\sum_n a_n e^{-z n} = z^{\alpha_1} e^{\kappa_1/z}, \\ 
&&\sum_n a_n (-1)^n e^{-z n} =  z^{\alpha_2} e^{\kappa_2/z},
\end{eqnarray}
with $\kappa_1>\kappa_2>0$. We can now consider the sum and difference of the two series, and upon rescaling $2z \to z$, we can apply the theorem. The final answer is then
\begin{eqnarray}
a_n \sim \left(\frac{n}{\kappa_1}\right)^{-\frac{\alpha_1}{2}-\frac{1}{2}} I_{-\alpha_1 -1}\left(2 \sqrt{n \kappa_1}\right)+ (-1)^n \left(\frac{n}{\kappa_2}\right)^{-\frac{\alpha_2}{2}-\frac{1}{2}} I_{-\alpha_2 -1}\left(2 \sqrt{n \kappa_2}\right)  + \cdots.
\end{eqnarray}
Which gives an asymptotic expansion for $a_n$. Here $I_\alpha(x)$ is the modified Bessel function of the first kind which arises after taking the inverse Laplace transform of $z^\alpha e^{\frac{\kappa}{\beta}}$. Going back to the problem of the number of partitions, covariance under different elements of the modular group give asymptotic expressions for the series $\sum_{n=0} p(n) \omega^n e^{-z n}$, where $\omega$ is a root of unity. This allows to write an asymptotic series for $p(n)$, where each term is written in terms of modified Bessel functions, as above. This was the result given by Hardy and Ramanujan. This series, however, is only asymptotic, and for any finite value of the spin it starts exploding at some point. As we review below, in the case of a modular invariant function (or covariant with appropriate weight) we can replace this asymptotic series by an exact expression.

\subsection{From asymptotic to convergent series}

\subsubsection{Modular transformations}
We will now focus our attention on functions 
\begin{equation}
Z(q) = \sum_{\mu>0} \frac{a_{-\mu}}{q^\mu} + \sum_{n=0}^\infty a_n q^n, 
\end{equation}
where $q=e^{2\pi i \tau}$. As a function of $\tau$ the function $Z(e^{2\pi i \tau})$ will be assumed to be holomorphic in the upper half plane with poles as $\tau \to i \infty$, as shown. We find it convenient to separate the principal part $\sum_{\mu>0} \frac{a_{-\mu}}{q^\mu}$ from the regular part as $\tau \to i \infty$. The modular group $PSL(2,\mathbb{Z})$ acts on $\tau$ by
\begin{equation}
\tau \to \frac{a \tau+ b}{s \tau -r},~~~~a r + b s = - 1,
\end{equation}
with integers $a,b,r,s$, positive $s$ and coprime $(r,s)=1$. We assume  $Z(q)$ transforms in a specific way (either invariantly or covariantly) under the entire group of modular transformations:
\begin{equation}
\label{modular1}
Z(e^{-\beta+ \frac{2\pi i r}{s}}) = f_{r,s}(\beta) Z(e^{-\frac{4\pi^2}{s^2 \beta} + \frac{2\pi i a}{s}}),
\end{equation}
where the weights $f_{r,s}(\beta)$ are given. For fixed $s$, the shifts $r \to r+s$ and $a \to a+s$ lead to exactly the same relations, hence we can take $0 < a,r \leq s$. In this region the relation above imposes a constraint whenever $a r + b s = - 1$ has integer solutions.

 Let us consider a few examples. For $s=1$ we only have $a=r=1$, and this fixes $b=-2$. For $s=2$ and $r=2$ the constraint $a r + b s = - 1$ does not have integer solutions. For $s=2,r=1$ it implies $a=1,b=-1$. In general, for $s=2,3,\cdots$, an integer solution exists for all $0<r<s$ provided $(r,s)=1$, {\it i.e. } $r,s$ are coprimes. In this case the Euler's totient theorem implies
\begin{equation}
a =a_{r,s} \equiv (-r)^{\phi(s)-1} \mod s,~~~~\phi(s) = \text{number of coprimes to $s$ between 1 and $s$},
\end{equation}
where $\phi(s)$ is the Euler's totient function. Let's now return to the relation (\ref{modular1}). Expressing both sides in their respective power expansions we obtain
\begin{equation}
\text{polar} + \sum_{n=0} a_n e^{\frac{2\pi i r}{s} n } e^{-\beta n} = f_{r,s}(\beta) \left(\sum_\mu a_{-\mu} e^{\mu \frac{4\pi^2}{s^2 \beta} +\mu  \frac{2\pi i a}{s}} + a_0 + \cdots \right).
\end{equation}
In the examples we consider $f_{r,s}(\beta)$ behaves as a power law as $\beta \to 0$, maybe up to an exponential term that can be absorbed by shifting $\mu$. Note that only the principal part on the r.h.s. contributes to the exponential behaviour. Following the discussion in the previous section we could write down an asymptotic expansion for $a_n$. This expansion is expected to be only asymptotic, and give a good approximation for large $n$. For a fixed, finite $n$, it is expected that the best estimate for $a_n$ is given by a finite number of terms, and after that adding new terms we would deviate from the correct result. Note furthermore that the use of the inverse Laplace transform, is not entirely justified: the transformations above, together with periodicity under $\beta \to \beta+2\pi i$,  imply the presence of an infinite tower of essential singularities along the real axis in the $\tau$ plane. 

As we will show below, modular invariance/covariance actually allows to do much better!  A powerful machinery was developed by Rademacher  \cite{Rademacher1} to give a {\it convergent series} for the number of partitions $p(n)$, introduced in the previous section. The method was further developed for other modular forms and further refined, see \cite{Rademacher2,Rademacher3}. It is a beautiful adaptation of the Hardy-Littlewood circle method, a technique of analytic number theory.  A detailed description of the method can be found in \cite{Rademacherbook}. Below we will discuss the main ideas of Rademacher's machinery and we will show its relation to the inverse Laplace transform. It turns out that the leading exponential asymptotic behaviour is unchanged, but each contribution contains in addition a tower of exponentially suppressed terms, with the net effect that the asymptotic series becomes now convergent.  

\subsubsection{Rademacher's circle method}
Consider again the series $Z(q) = \text{principal}+\sum_{n=0} a_n q^n$. The coefficients $a_n$ are given by the Cauchy residue theorem
 \begin{equation}
 a_n = \oint \frac{dq}{q} \frac{Z(q)}{q^n},
 \end{equation}
where the contour encloses the origin and has radius smaller than 1. In the $\tau$ plane the integral becomes
 \begin{equation}
 a_n = \int_{\Gamma}  Z(q) e^{-2\pi i \tau n } d\tau,
 \end{equation}
where the contour ${\Gamma}$ is chosen to be a path between $\tau=i$ and $1+i$.  The next step is to use modular covariance 
\begin{equation}
Z(e^{-\beta+ \frac{2\pi i r}{s}}) = f_{r,s}(\beta) \left( a_{-\mu} e^{\mu \frac{4\pi^2}{s^2 \beta} -\mu  \frac{2\pi i a}{s}} + \cdots \right).
\end{equation}
For simplicity we will assume $f_{r,s}(\beta) \sim \beta^\alpha$ and take into account the contribution of a single polar term. All other polar terms can be treated in exactly the same way, while  regular terms are exponentially suppressed and will not contribute to the Rademacher expansion.
This modular transformation implies that in the $\tau-$plane $Z(e^{2\pi i \tau})$  has an essential singularity at each point of the form $r/s$ with integer $r,s$. What Rademacher did was to split the contour ${\Gamma}$ in a series of smaller contours ${\Gamma}_{r/s}$ such that in each of those only the essential singularity at $r/s$ contributes. Let us now explain this construction. 

\bigskip

\noindent{\bf Farey sequences and Ford circles}

\bigskip

\begin{figure}[h]
  \centering
\includegraphics[width=130mm]{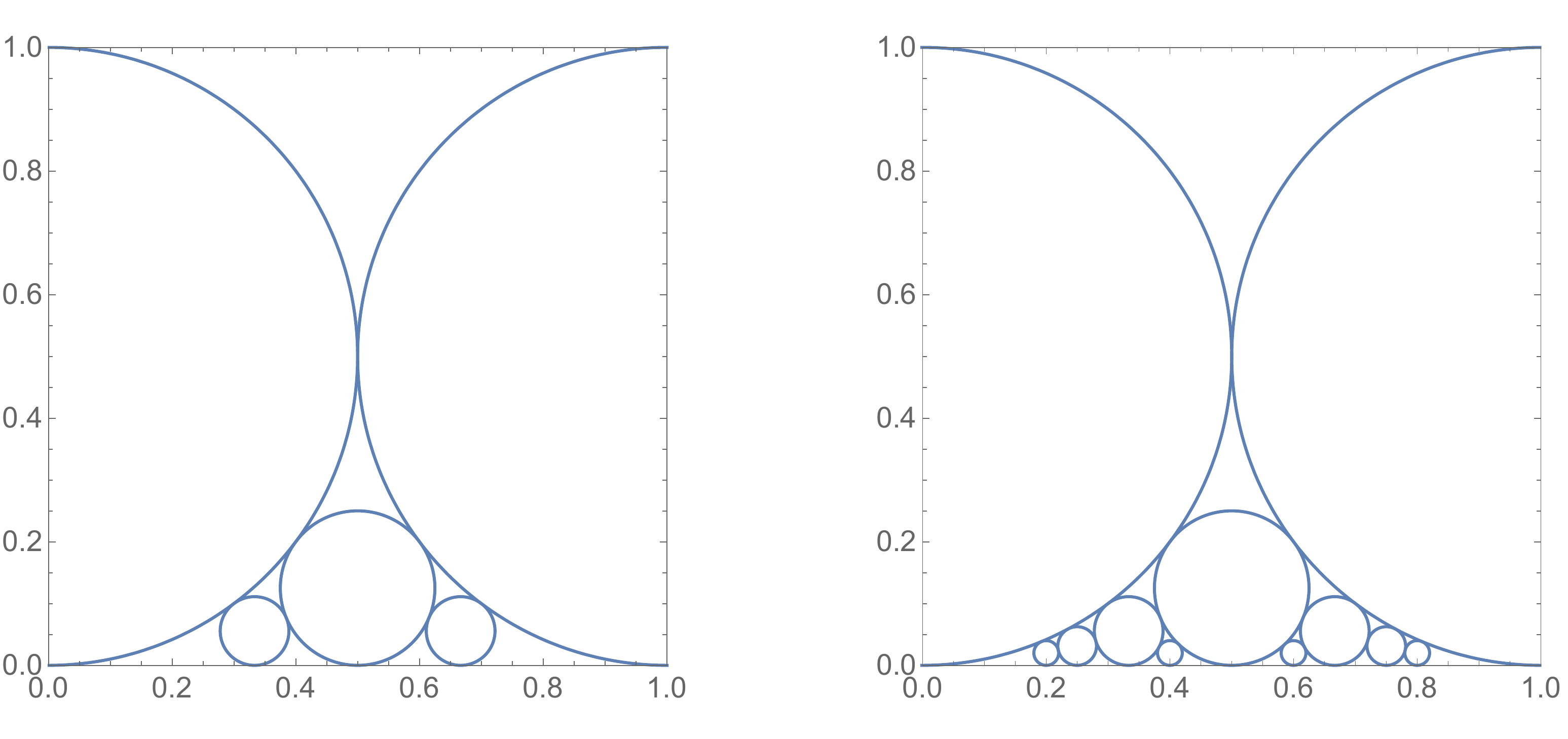}
  \caption{Ford circles corresponding to the Farey sequences of order three (left) and five (right)}
  \label{fordcircles}
\end{figure}

\noindent Consider the segment $[0,1]$ and mark the initial and final points which we write as $0/1$ and $1/1$. This is the sequence at order one. Now mark all rational points of the form $h/2$ that were not marked before, namely only $1/2$. All marked points give the sequence at order two. Now mark all rational points of the form $h/3$ that were not marked before, namely $1/3$ and $2/3$. This gives the sequence at order three, and so on.  For example the Farey sequence at order 5 is given by the following marked points
\begin{equation}
\left\{0,\frac{1}{5},\frac{1}{4},\frac{1}{3},\frac{2}{5},\frac{1}{2},\frac{3}{5},\frac{2}{3},\frac{3}{4},\frac{4}{5},1\right\}.
\end{equation}
Next, given a Farey sequence of a given order, we draw a series of circles in the $\tau-$plane. To each fraction $\frac{r}{s}$ with $(r,s)=1$ we associate a circle $C_{rs}$ with centre $\tau_{rs} = \frac{r}{s}+ \frac{i}{2s^2}$ and radius $\frac{1}{2s^2}$. In fig. \ref{fordcircles} we can see the circles corresponding to the Farey sequences of third and fifth order.
Note that each circle touches the real axis, and that given two consecutive fractions $r/s$ and $r'/s'$ in the Farey sequence, their respective circles touch each other, but do not intersect. 

\bigskip

\noindent{\bf Rademacher's contour}

\bigskip

\begin{figure}[h]
  \centering
\includegraphics[width=60mm]{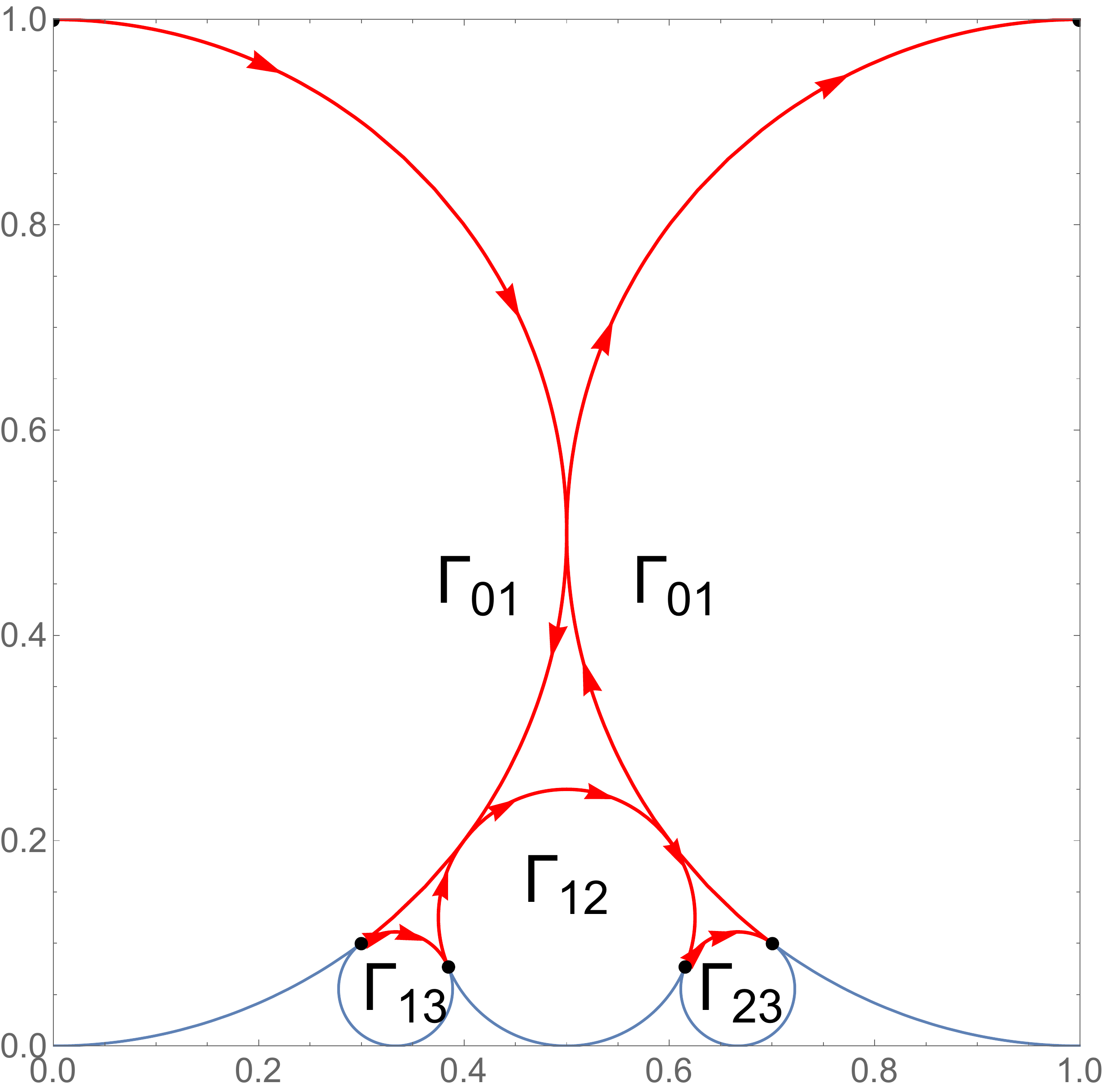}
  \caption{For $N=3$ we deform the contour as shown in the figure and write $\Gamma = \Gamma_{01} \cup \Gamma_{13} \cup \Gamma_{12} \cup \Gamma_{23} \cup \Gamma_{11}$.}
  \label{RademacherContour}
\end{figure}

\noindent Let's go back to the problem of computing $a_n$. Above we have written them in terms of an integral over the $\tau$ plane along a path between $\tau=i$ and $\tau=i+1$. The simplest contour would be a straight segment between these two points. Rademacher instead, considered the following sequence of contours. Given a Farey sequence of order $N$, we defined the contour as follows.  Start from the point $\tau=i$ and follow the Ford circle $C_{01}$. We follow this circle until it touches with the consecutive circle, which for the case at hand will be $C_{1N}$. Now we follow this circle until it teaches the consecutive circle, and so on. For instance, fig. \ref{RademacherContour} shows the Rademacher's contour for the case $N=3$. 
The crucial observation by Rademacher was the following: this construction naturally splits the path of integration $\Gamma$ into a series of consecutive paths $\Gamma_{rs}$. For forms of non-positive weight it turns out that for each path $\Gamma_{rs}$ only the corresponding essential singularity at $\tau = r/s$ contributes, and all the others can be ignored. The error in doing this tends to zero as we take $N \to \infty$. Note that for each finite $N$, the paths never touch the real line, so that the essential singularities are avoided. 

\bigskip

\noindent{\bf Computing the integrals and final expression}

\bigskip

\begin{figure}[h]
  \centering
\includegraphics[width=65mm]{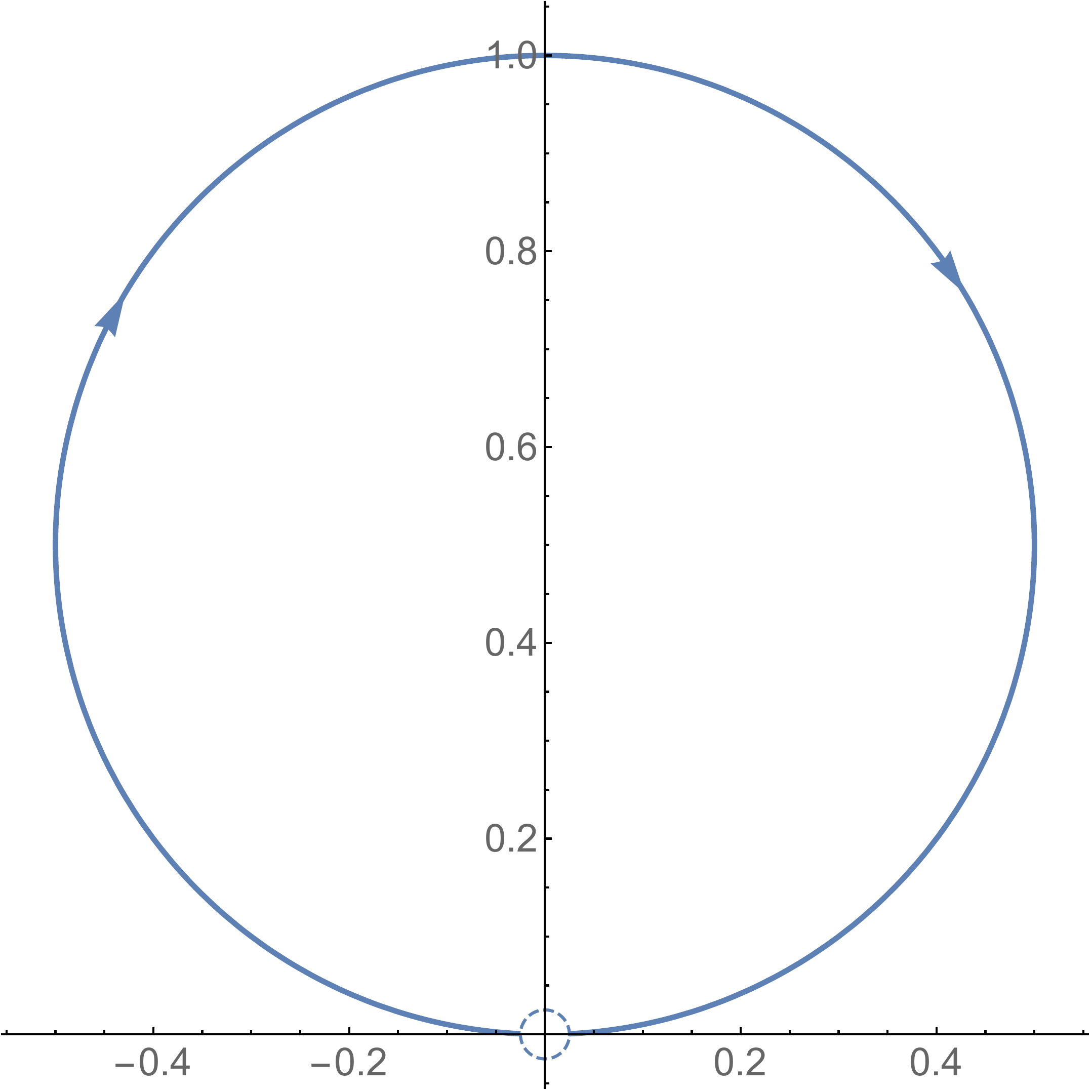}
  \caption{$\Gamma_{01} \cup \Gamma_{01}$ tend to the circle  $C'_{01}$, shown in the figure, as $N \to \infty$.}
  \label{C01}
\end{figure}

\noindent Let us analyse the leading contribution, corresponding to $s=1$ and arising from the singularity at $0$ (and at $1$ by periodicity). The corresponding path is $\Gamma_{s=1}  = \Gamma_{01} \cup \Gamma_{11}$. By periodicity we can shift $\Gamma_{11}$ the the left, so that the path $\Gamma_{s=1}$ is just an arc. As $N \to \infty$ the path tends to the circle   $C_{01}$ with centre $\tau =\frac{i}{2}$ and radius $1/2$, where the origin is excluded, see fig. \ref{C01}.
Let's denote this path by $C'_{01}$ to stress the fact that the origin is excluded. We are then led to compute the following integral
\begin{equation}
a_n^{s=1} = \int_{C_{01}'} \beta^\alpha e^{\frac{\kappa}{\beta}} e^{-2\pi i \tau n } d\tau,
\end{equation}
where recall $\beta = -2\pi i \tau$ and we have introduced $\kappa = 4\pi^2 \mu$. We now make a change of variables, and go from $\tau$ to $w = \frac{1}{\beta} = -\frac{1}{2 \pi i \tau}$. The resulting integral can then be evaluated in terms of the  Bessel function of the first kind 
\begin{equation}
a_n^{s=1} = \frac{1}{2\pi i} \int_{\frac{1}{2\pi} - i \infty}^{\frac{1}{2\pi} + i \infty} w^{-\alpha} e^{\kappa w} e^{\frac{n}{w}} \frac{dw}{w^2} = \left( \frac{\kappa}{n} \right)^\frac{1+\alpha}{2} I_{\alpha +1}\left(2 \sqrt{\kappa  n}\right). 
\end{equation}
For latter convenient we introduce a Rademacher inversion function ${\cal R}^{-1}\left(f(\beta) \right)$, defined by the Rademacher integrals, and such that 
\begin{equation}
{\cal R}^{-1}\left(\beta^\alpha e^{\frac{\kappa}{\beta}} \right)=\left( \frac{\kappa}{n} \right)^\frac{1+\alpha}{2} I_{\alpha +1}\left(2 \sqrt{\kappa  n}\right).
\end{equation}
We can proceed in exactly the same way for every essential singularity. In case of a general modular transformation  (\ref{modular1}) of exponential form, the final result for the coefficients $a_n$ takes the form
\begin{eqnarray}
a_n  = \sum_{\substack{s=1,2,\cdots}} \sum_{\substack{0<r \leq s,\\(r,s)=1}}  e^{\frac{-2\pi i r}{s} n }  {\cal R}^{-1}\left\{ f_{r,s}(\beta) \sum_{\mu>0} a_{-\mu} e^{\mu \frac{4\pi^2}{s^2 \beta} -\mu  \frac{2\pi i a}{s}}  \right\}. \nonumber
\end{eqnarray}
Note that only the principal part on the r.h.s. will contribute to this series. The claim is then that for a large class of modular forms the above series converges to the correct $a_n$, allowing to reconstruct the full modular form from its principal part. Note that in order for this to be true the sum over $s$ must be convergent, and the error terms from the Rademacher procedure must tend to zero as $N \to \infty$. Rademacher showed that this was the case for the generating function of partitions in his original work \cite{Rademacher1}, while later with Zuckerman and by himself \cite{ZuckermanRademacher,Rademacher2} he extended the treatment to modular forms of non-positive weight. The status for positive weights is more complicated, see for instance \cite{Cheng:2012qc}. For weights higher than two, the Rademacher procedure converges to the right result, while for positive weights smaller than two  the Rademacher procedure may converge, but not necessarily to the correct result. In this range the Rademacher construction will generally lead to a mock modular form, as opposed to a modular form. 

Let us comment on the difference between this result and a naive inverse Laplace transform. The relevant integrand is the same in both cases, but the contours are different. The inverse Laplace transform gives
\begin{eqnarray}
{\cal L}^{-1}\left(\beta^\alpha e^{\frac{\kappa}{\beta}} \right) = \frac{1}{2\pi i} \int_{\sigma-i \infty}^{\sigma+ i \infty} \beta^\alpha e^{\frac{\kappa}{\beta}} e^{\beta n} d\beta = \left( \frac{\kappa}{n} \right)^\frac{1+\alpha}{2} I_{-\alpha -1}\left(2 \sqrt{\kappa  n}\right).
\end{eqnarray}
 The results are very similar, but the sign on the Bessel function has changed. The asymptotic behaviour for the Bessel function for $x$ real implies
\begin{equation}
I_\nu(x) - I_{-\nu}(x) =e^{-x}\left( -\sqrt{\frac{2}{\pi }} \sqrt{\frac{1}{x}} \sin (\pi  \nu )+ \cdots \right),
\end{equation}
so that the difference between the two results is exponentially small for large $x$. This small difference, however, makes the Rademacher expansion convergent for finite $n$. 

\subsection{Examples}
\label{examples}
\bigskip

\noindent{\bf Example 1}

\bigskip

\noindent Let us now consider the $j-$invariant function $J(q) = j(q)-744$. This is a modular invariant function with the Fourier expansion 
\begin{equation}
J(q) = \sum_{n=-1} a_n q^n = \frac{1}{q} + 196884 q + \cdots.
\end{equation}
Since $J(q)$ is modular invariant, we simply get
\begin{equation}
J\left( e^{-\beta + \frac{2\pi i r}{s}} \right) = e^{\frac{4\pi^2}{s^2\beta} - \frac{2\pi i a}{s}} + \cdots.
\end{equation}
So that the Fourier coefficients are given by the following Rademacher sum
\begin{eqnarray}
a_n  &=& \sum_{\substack{s=1,2,\cdots}} \sum_{\substack{0<r \leq s,\\(r,s)=1}} 2\pi  e^{-\frac{2\pi i r}{s} n - \frac{2\pi i a(r,s)}{s}} \frac{I_1\left(\frac{4 \sqrt{n} \pi }{s}\right)}{\sqrt{n} s} \nonumber \\
&=& \sum_{\substack{s=1,2,\cdots}} K(-n,1,s) \frac{2\pi}{\sqrt{n} s}I_1\left(\frac{4 \sqrt{n} \pi }{s}\right),
\end{eqnarray}
where we have introduced the Kloosterman sum 
 \begin{equation}
K(a,b;s) = \sum_{\substack{0<r \leq s,\\(r,s)=1}} e^{\frac{2 \pi i}{s} (a r+b r^{-1})}.
\end{equation}
Here $r^{-1}$ is the inverse of $r$ modulo $s$, namely an integer such that $r r^{-1} = 1 \mod s$. The convergence of this expression depends on estimates for the Kloosterman sum at large $s$. The Weil bound implies
\begin{equation}
K(a,b;s)  \leq \tau(s) \sqrt{(a,b,s)} \sqrt{s},
\end{equation}
for $a,b \neq 0$, where $(a,b,s)$ is the greatest common divisor of $a,b$ and $s$, and $\tau(s)$ is the number of positive divisors of $s$, which grows at most logarithmically with $s$. On the other hand we have 
\begin{equation}
\frac{2\pi}{\sqrt{n} s} I_1\left(\frac{4 \sqrt{n} \pi }{s}\right) \sim \frac{1}{s^2}.
\end{equation}
So that for fixed $n \neq 0$ the sum is convergent. On the other hand, for $n=0$ we have $K(0,1,s) \sim s$ (for instance, for prime numbers), so that the sum naively fails to converge. This can also be understood as follows: To $J(q)$ we can always add a constant, which is certainly modular invariant. This does not contribute to its principal part but it changes the coefficient $a_0$. It turns out that the Rademacher expansion selects specific constant, see \cite{Duncan:2009sq}.\footnote{We thank Alex Maloney for drawing \cite{Duncan:2009sq} to our attention.}

\bigskip

\noindent{\bf Example 2}

\bigskip

\noindent Let us consider again the generating function for partitions of $n$. 
\begin{equation}
Z(q) = \sum_{n=0} a_n q^n = \frac{q^{\frac{1}{24}}}{\eta(\tau)} = 1+ q+ 2 q^2+ 3 q^3+\cdots.
\end{equation}
The Dedekind eta function has the following transformation rules under modular transformations
\begin{equation}
\eta \left(\frac{a \tau+b}{s \tau-r} \right) = \epsilon(r,s) (s \tau-r)^{1/2} \eta(\tau),
\end{equation}
where for $s>0$
\begin{equation}
\epsilon(r,s) = e^{i \pi \left(\frac{a-r}{12 s }+ S[r,s]- \frac{1}{4} \right)},
\end{equation}
with $S[r,s]$ the Dedekind sum
\begin{equation}
S(r,s) = \sum_{n=1}^{s-1} \frac{n}{s}\left( \frac{r n}{s} - \left[ \frac{r n}{s}\right] -\frac{1}{2} \right),~~~S[r,1]=0.
\end{equation}
With this transformation at hand we find
\begin{equation}
Z(q) =  \frac{q^{\frac{1}{24}}}{\eta(\tau)} = \frac{q^{\frac{1}{24}}}{q'^{\frac{1}{24}}} \epsilon(r,s)(s \tau-r)^{1/2} Z(q'),
\end{equation}
so that
\begin{equation}
Z\left( e^{-\beta + \frac{2\pi i r}{s}} \right) = \sqrt{\frac{\beta s}{2\pi}} e^{-\frac{\beta}{24}} e^{\frac{\pi^2}{6 s^2 \beta}} e^{i \pi S[r,s]} \left(1 + \cdots\right),
\end{equation}
where the next terms after the 1 are exponentially suppressed and do not contribute to the Rademacher expansion.  Applying the result above we obtain
\begin{eqnarray}
a_n  &=& \sum_{\substack{s=1,2,\cdots}} \sum_{\substack{0<r \leq s,\\(r,s)=1}} e^{-\frac{2\pi i r}{s} n }  e^{i \pi S[r,s]}  {\cal R}^{-1}\left\{ \sqrt{\frac{\beta s}{2\pi}} e^{-\frac{\beta}{24}} e^{\frac{\pi^2}{6 s^2 \beta}}  \right\} \nonumber\\
&=& \sum_{\substack{s=1,2,\cdots}} \sum_{\substack{0<r \leq s,\\(r,s)=1}}  \sqrt{s} e^{-\frac{2\pi i r}{s} n }  e^{i \pi S[r,s]}  \partial_n \frac{2 \sqrt{3} \sinh \left(\frac{\pi  \sqrt{24 n-1}}{6 s}\right)}{\pi  \sqrt{24 n-1}}.
\end{eqnarray}
Similar estimates to the ones for the Kloosterman sum exist in the case of the extra insertion of $S[r,s]$ in the exponent, see \cite{Kloosterman}. On the other hand 
\begin{equation}
\partial_n \frac{2 \sqrt{3} \sinh \left(\frac{\pi  \sqrt{24 n-1}}{6 s}\right)}{\pi  \sqrt{24 n-1}} \sim \frac{1}{s^3},
\end{equation}
so that the sum over $s$ is convergent, and actually converges to the right result for $n \geq 0$. 

\bigskip

\noindent{\bf Example 3}

\bigskip
For positive and small weight, even if convergent, there is no warrantee the Rademacher sum will converge to the correct result. Let us see an example of this. Consider the product
\begin{equation}
Z(q) =  q^{-\frac{1}{24}} \eta(\tau)J(q) = \frac{1}{q} - 1 + 196883 q + \cdots.
\end{equation} 
This is relevant when decomposing the partition function of the Monster CFT in Virasoro characters. In this case the Rademacher sum gives
\begin{eqnarray}
a_n  &=& \sum_{\substack{s=1,2,\cdots}} \sum_{\substack{0<r \leq s,\\(r,s)=1}} e^{-\frac{2\pi i r}{s} n }  e^{-i \pi S[r,s]}  {\cal R}^{-1}\left\{ \sqrt{\frac{2\pi}{\beta s}} e^{\frac{\beta}{24}} e^{\frac{23 \pi^2}{6 s^2 \beta}- \frac{2\pi i a(r,s)}{s}}  \right\} \nonumber\\
&=& \sum_{\substack{s=1,2,\cdots}} \sum_{\substack{0<r \leq s,\\(r,s)=1}} e^{-\frac{2\pi i r}{s} n -i \pi S[r,s]- \frac{2\pi i a(r,s)}{s}} \frac{\sqrt{2} \sinh \left(\frac{\sqrt{\frac{46}{3}} \pi  \sqrt{n+\frac{1}{24}}}{s}\right)}{\sqrt{\left(n+\frac{1}{24}\right) s}}.
\end{eqnarray}
We have checked numerically that this converges to the correct result, although super slowly, up to a piece proportional to\footnote{Numerically very close to $36q^{-\frac{1}{24}} \eta(\tau)$.} $q^{-\frac{1}{24}} \eta(\tau)$. Indeed, $q^{-\frac{1}{24}} \eta(\tau)$ has exactly the same weight as $Z(q)$ but no polar part, so we always have the freedom of adding it. A similar phenomenon occurs when considering N copies of the Monster CFT. 

\section{Application to 2d CFT}
\label{Rademacher2dCFT}
Consider a unitary 2d CFT with central charge $c>1$. Its states are classified by their conformal weights $(h,\bar h)$ or alternatively their dimension and spin
\begin{equation}
\Delta = h + \bar h,~~~ j = h - \bar h.
\end{equation}
We will consider the partition function of the theory on a torus with complex structure moduli $q=e^{2\pi i \tau}$ and $\bar q=e^{-2\pi i \bar \tau}$. We will assume the CFT possesses Virasoro symmetry but not an extended chiral algebra. The partition function can be expanded in Virasoro characters and takes the form
\begin{equation}
Z(q,\bar q) = \chi_0(q) \chi_0(\bar q) + \sum_{h,\bar h >0} n_{h,\bar h}  \chi_h(q) \chi_{\bar h}(\bar q) +\sum_{h>0} N_{h}  \chi_h(q) \chi_{0}(\bar q)+\sum_{\bar h >0} N_{\bar h}  \chi_0(q) \chi_{\bar h}(\bar q).
\end{equation}
The term $ \chi_0(q) \chi_0(\bar q)$ corresponds to the vacuum, the terms $\chi_h(q) \chi_{0}(\bar q),\chi_0(q) \chi_{\bar h}(\bar q)$ correspond to conserved currents, while $\chi_h(q) \chi_{\bar h}(\bar q)$ corresponds to non-degenerate states. Here $n_{h,\bar h}$, etc, denote the multiplicity with which such operators appear. For a unitary CFT they are non-negative integer numbers. The Virasoro characters are given by
\begin{equation}
\chi_0(q) = (1-q) \frac{q^{-\frac{c-1}{24}}}{\eta(q)},~~~~\chi_{h>0}(q) =  \frac{q^{h-\frac{c-1}{24}}}{\eta(q)}.
\end{equation}
It will be convenient to introduce the concept of twist, given by $\tau=min(h,\bar h)$. Although not strictly necessary, we will assume the theory has a twist gap $\tau_{gap}$. This forbids the presence of conserved currents. We expect this to be the generic situation for irrational CFTs with $c>1$, although no such explicit examples are known. In this case the partition function takes the form  
\begin{equation}
Z(q,\bar q) = \chi_0(q) \chi_0(\bar q) + \sum_{h,\bar h \geq \tau_{gap}}  \chi_h(q) \chi_{\bar h}(\bar q),
\end{equation}
where we have left implicit the multiplicities $n_{h,\bar h}$. Note that for a generic irrational CFTs with no extra symmetries, we expect most multiplicities to be 1. A crucial property of the partition function is modular invariance. $PSL(2,\mathbb{Z})$ acts on the complex moduli of the torus $\tau,\bar \tau$ as follows
\begin{equation}
\tau \to \tau= \frac{a \tau +b}{s \tau-r},~~~~~\bar \tau \to \bar \tau= \frac{a \bar \tau +b}{s \bar \tau-r},
\end{equation}
for integers $a,b,s,r$ with positive $s$ and $a r+b s=-1$, and in particular $(r,s)=1$. This maps the torus to an equivalent one, and hence the corresponding partition functions should agree. This implies
\begin{equation}
Z\left( q,\bar q \right) = Z\left(q',\bar q'\right),
\end{equation}
where 
\begin{eqnarray}
&q = e^{-\beta+ \frac{2\pi i r}{s}},~~~~~ q' = e^{-\frac{4\pi^2}{s^2 \beta} + \frac{2\pi i a}{s}}, \\ 
&\bar q = e^{-\bar \beta- \frac{2\pi i r}{s}},~~~~~\bar q' = e^{-\frac{4\pi^2}{s^2 \bar \beta} - \frac{2\pi i a}{s}}.
\end{eqnarray}
We will analyse the constraints on the spectrum imposed by modular invariance and the twist gap. In particular, we will follow and revisit \cite{Benjamin:2019stq} in view of our discussion in the previous section. As in \cite{Benjamin:2019stq} it will be important to consider $\beta,\bar \beta$ as independent variables, however, for us it will be important to consider them complex, with positive real part. As explained in \cite{Mukhametzhanov:2019pzy} this is possible thanks to unitarity. 

\subsection{Constraints on the spectrum}
\begin{figure}[t!]
  \centering
\includegraphics[width=130mm]{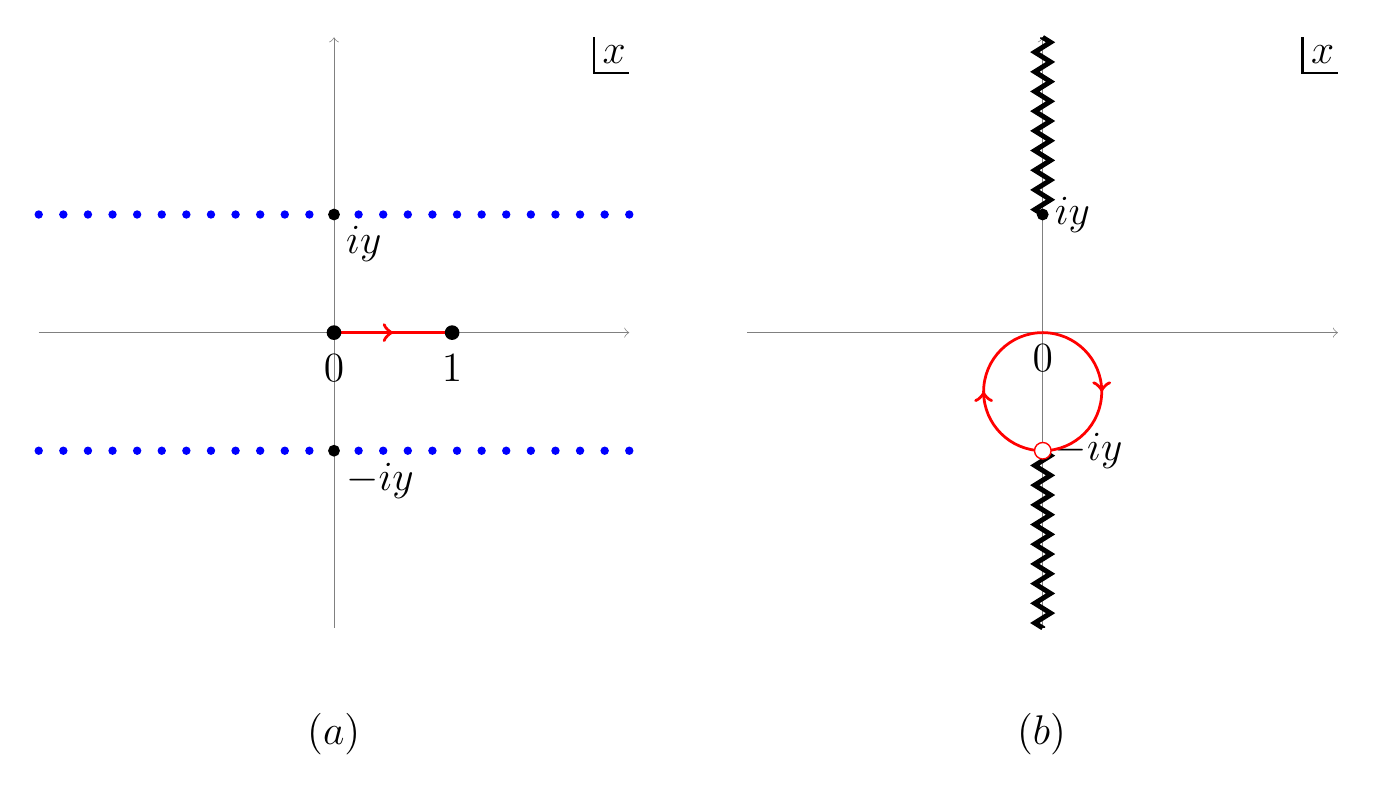}
  \caption{On the left we see the structure of essential singularities in the $x-$plane, at all points of the form $x = \pm i y + \mathbb{Q}$. On the right we have performed a Rademacher deformation as to isolate only two of those, at $x = \pm i y$.}
  \label{Rademacherdef}
\end{figure}

\begin{figure}[t!]
  \centering
\includegraphics[width=130mm]{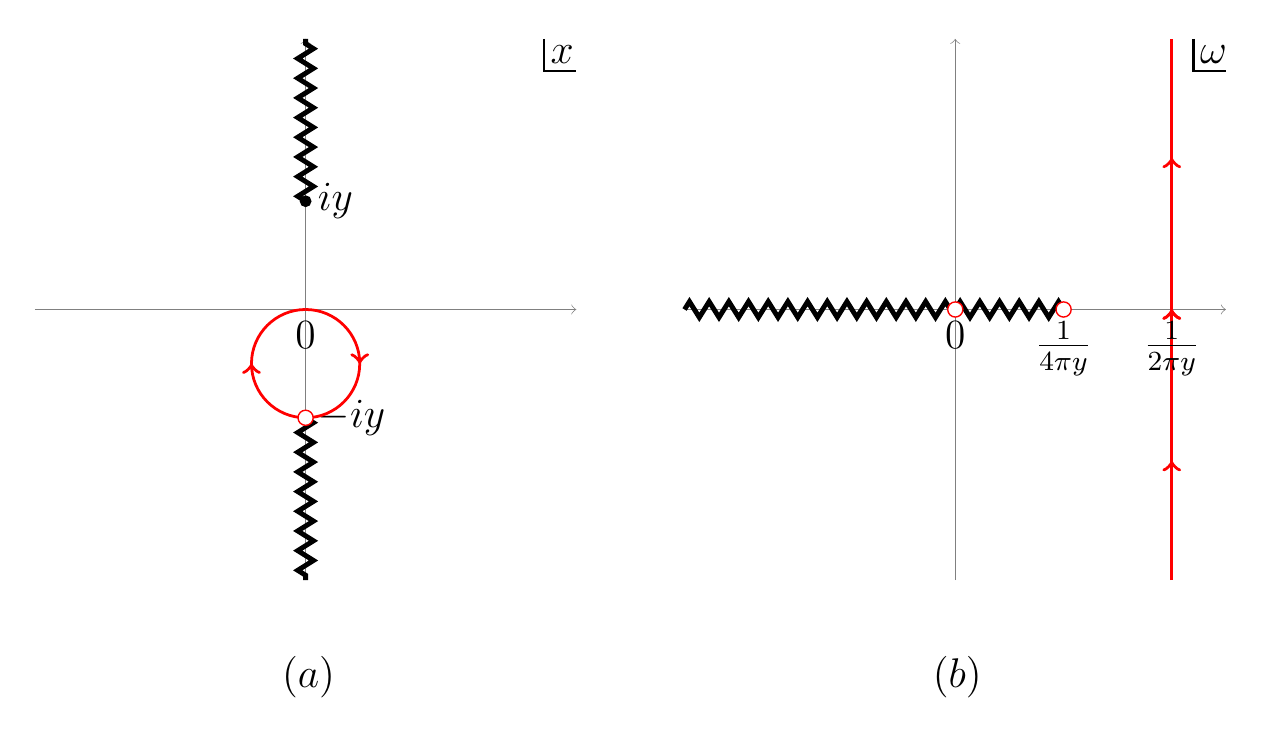}
  \caption{The Rademacher contour for a single pole (left) is mapped to a straight in the $w-$plane (right). }
  \label{wplane}
\end{figure}

Following \cite{Keller:2014xba} we define the partition function $Z^p(q,\bar q) = y^{1/2} \eta(q) \eta(\bar q) Z(q,\bar q) $ where $y = \operatorname{Im}(\tau)$. This is clearly modular invariant and has the decomposition 
\begin{equation}
Z^p(q,\bar q)  = y^{1/2} \left( q^{-\hat c}\bar q^{-\hat c} (1-q)(1-\bar q) + \sum_{h,\bar h} q^{h-\hat c} \bar q^{\bar h-\hat c} \right),
\end{equation}
where we have introduced a short-hand notation for $\frac{c-1}{24} = \hat c$, the sum runs over Virasoro primaries and we have singled out the contribution of the vacuum. Modular invariance $Z^p(q,\bar q) = Z^p(q',\bar q')$ together with this decomposition leads to
\begin{equation}
\label{modular2d}
vac+  \sum_{h,\bar h} q^{h-\hat c} \bar q^{\bar h-\hat c} =  \sqrt{\frac{y'}{y}}\left( vac' + \sum_{h,\bar h} q'^{h-\hat c} \bar q'^{\bar h-\hat c} \right),
\end{equation}
where we have single-out the contribution of the vacuum on both sides. For a general modular transformation we have 

\begin{equation}
\tau = x+i y = i \frac{\beta}{2\pi}+\frac{r}{s},~~~\bar \tau = x-i y = -i \frac{\bar \beta}{2\pi}+\frac{r}{s}
\end{equation}
so that $y = \frac{\beta+\bar \beta}{4\pi}$, while $y' = \frac{\pi}{s^2} \left(\frac{1}{\beta} + \frac{1}{\bar \beta} \right)$, which leads to 
\begin{equation}
\sqrt{\frac{y'}{y}} =\frac{1}{s}\sqrt{ \frac{4\pi^2}{\beta \bar \beta}}.
\end{equation}
We will now consider the following problem. Given an isolated operator of weights $(h_0,\bar h_0)$, such that $h_0+\bar h_0 < 2\hat c$,  in the r.h.s of (\ref{modular2d}) what are the implications for the density of operators (to be defined below) on the left?  The modular constraints in this case give

\begin{equation}
vac+  \sum_{h,\bar h} e^{(-\beta+ \frac{2\pi i r}{s})(h-\hat c)} e^{(-\bar \beta- \frac{2\pi i r}{s})(\bar h-\hat c)}  =  \frac{1}{s}\sqrt{ \frac{4\pi^2}{\beta \bar \beta}}  e^{(-\frac{4\pi^2}{s^2 \beta} + \frac{2\pi i a}{s})(h_0-\hat c)}e^{(-\frac{4\pi^2}{s^2 \bar \beta} - \frac{2\pi i a}{s})(\bar h_0-\hat c)}.
\end{equation}
Note that there is one such relation for each pair of positive coprimes $(r,s)$. We find it convenient to rewrite these relations in terms of $x,y$ and the dimensions and spins of the operators\footnote{This equation may look confusing at first, because the l.h.s. does not contain $r,s$ while the r.h.s. does. The l.h.s. represents the full partition function and hence is modular invariant. The r.h.s represents the contribution of a single operator, with all others left implicit, and hence is not. }
\begin{equation}
F(x,y)\equiv \sum_{j,\Delta} e^{2\pi i j x} e^{-2\pi y e}= \frac{1}{s} \sqrt{\frac{1}{(x-r/s)^2+y^2}} e^{\frac{2\pi i a}{s} j} e^{\frac{2 i \pi  (r-s x)}{s (r-s x)^2+s^3 y^2} J} e^{-\frac{2 \pi  y E}{(r-s x)^2+s^2 y^2}},
\end{equation}
where $J$ denotes the spin of the operator on the r.h.s and, following \cite{Maloney:2007ud,Keller:2014xba}, we have introduced the shifted dimensions 
\begin{equation}
e= \Delta-2 \hat c,~~~~E=\Delta_0 - 2\hat c,
\end{equation}
with $\Delta_0 = h_0 + \bar h_0$. As already mentioned, we will consider $\beta,\bar \beta$ as independent complex variables. This means we consider $x,y$ as independent complex variables as well. We would like to understand the constraints on the spectrum from the presence of the operator on the r.h.s. of these relations. In order to do this, we will apply the ideas of the previous section. the first difference is the presence of two quantum numbers: $\Delta$, or equivalently $e$, and $j$. The sum over $j$ is discrete, while the spectral density, for a given spin, will turn out to be continuous. We rewrite the l.h.s. as follows
\begin{equation}
\sum_{j,\Delta} e^{2\pi i j x} e^{-2\pi y e}= \sum_{j=-\infty}^\infty a_j(y)e^{2\pi i j x},
\end{equation}
with $a_j(y) = \sum_{\Delta}e^{-2\pi y e}$ where the sum runs over operators with a given spin $j$. 
\begin{equation}
\sum_{j=-\infty}^\infty a_j(y)e^{2\pi i j x}= \frac{1}{s} \sqrt{\frac{1}{(x-r/s)^2+y^2}} e^{\frac{2\pi i a}{s} j} e^{\frac{2 i \pi  (r-s x)}{s (r-s x)^2+s^3 y^2} J} e^{-\frac{2 \pi  y E}{(r-s x)^2+s^2 y^2}}.
\end{equation}
The second difference is that now the sum runs over positive and negative spins, we will return to this point later. From standard Fourier theory 
\begin{equation}
a_j(y) = \int_0^1 F(x,y) e^{-2\pi i j x}dx.
\end{equation}
Let's consider the integrand in the $x-$plane. The modular relations imply the presence of a tower of essential singularities at points
\begin{equation}
x = \pm i y + \frac{r}{s}.
\end{equation}
$r/s$, together with periodicity $x \to x+1$ cover all rational points, so there is an essential singularity at any point of the form $x = \pm i y + \mathbb{Q}$. Following the procedure sketched in the previous section, we can deform the path so that the contribution of each essential singularity can be taken into account independently. Let us focus in the one corresponding to $r=0,s=1$. All others will work in very much the same way. The procedure below will not work for $j=0$. For $j \neq 0$, depending on the sign of $j$ we need to deform the contour upwards or downwards. For definiteness we assume $j>0$. In this case the contour must be deformed downwards, towards the singularity at $x=-i y$. The integral to be computed is then
\begin{equation}
a^{s=1}_j(y) = \int_{C'_{0,1}} \frac{e^{-\frac{2\pi i x}{x^2+y^2} J} e^{-\frac{2\pi y}{x^2+y^2}E}}{\sqrt{x^2+y^2}} e^{-2\pi i j x}dx,
\end{equation}
where the path $C'_{0,1}$ is a circle with centre at $x=-i y/2$ and radius $y/2$, and the point $x=-i y$ is excluded, as in our discussion in the previous section, and we have taken $y$ as real and positive. Note that the integrand has two branch cuts $x \in (-i y,-i \infty)$ and $x \in (i y,i \infty)$. See fig. \ref{Rademacherdef}.  
We now make the change of variables $x \to w =-\frac{1}{2 \pi  i (x+i y)}$. This maps the branch cut  $(-i y,-i \infty)$ in the $x-$plane to the branch cut $(-\infty,0)$ and $(i y,i \infty)$ to $(0,\frac{1}{4\pi y})$. In this variables the Rademacher path is a straight line, with real part $w= \frac{1}{2\pi y}$, see fig. \ref{wplane}.
In this variables the integral becomes

\begin{equation}
\label{Rademachersingle}
a^{s=1}_j(y) = -i \int_{\frac{1}{2\pi y}-i \infty}^{\frac{1}{2\pi y}+i \infty} \frac{2 e^{j(\frac{1}{w}-2\pi y)}}{w \sqrt{-1+4 \pi w y}} e^{\frac{4 \pi ^2 w (1-2 \pi  w y)}{4 \pi  w y-1} J} e^{-\frac{8 \pi ^3 w^2 y}{4 \pi  w y-1}E} dw.
\end{equation}
It is instructive to look at the exponential factor for large values of $w$

\begin{equation}
e^{\frac{4 \pi ^2 w (1-2 \pi  w y)}{4 \pi  w y-1} J} e^{-\frac{8 \pi ^3 w^2 y}{4 \pi  w y-1}E} \sim e^{-2\pi^2 (E+J) w}.
\end{equation}
This suggests a way to solve the integral. Introducing $\kappa_{\pm} = -(E \pm J)$ we can rewrite the integral as 

\begin{equation}
a^{s=1}_j(y) = -i \int_{\frac{1}{2\pi y}-i \infty}^{\frac{1}{2\pi y}+i \infty} \frac{2 e^{j(\frac{1}{w}-2\pi y)}}{w \sqrt{-1+4 \pi w y}} e^{-\frac{2 \pi ^2 w}{1-4 \pi  w y} \kappa_-} e^{2\pi^2 \kappa_+ w } dw.
\end{equation}
But up to an overall factor, this is nothing but the definition of the Inverse Laplace transform with dual variable $2\pi^2 \kappa_+$. Expanding in powers of $j$ and $\kappa_-$ we can invert term by term using the following identity
\begin{equation}
\frac{1}{2\pi i} \int_{\frac{1}{2\pi y}-i \infty}^{\frac{1}{2\pi y}+i \infty} \frac{w^\alpha}{(4\pi w y-1)^\beta} e^{2\pi^2 \kappa_+ w } dw = \frac{ (-\kappa_+)^{-\alpha +\beta -1} \, _1F_1\left(\beta ;\beta -\alpha ;\frac{\kappa_+ \pi }{2 y}\right)}{2^{\alpha+\beta+1}  \pi^{2 \alpha -\beta +2} y^\beta \Gamma (\beta -\alpha )}.
\end{equation}
We obtain
\begin{eqnarray}
a^{s=1}_j(y) &=& \sum_{p,q=0}e^{-2\pi y j} \frac{j^p \kappa_+^{p+\frac{1}{2}} 2^{3 p-q+\frac{3}{2}} \pi ^{2 p+q+1} \kappa_-^q  \, _1F_1\left(q+\frac{1}{2};p+\frac{3}{2};\frac{\pi \kappa_+}{2 y}\right)}{y^{q+\frac{1}{2}} \Gamma (2 p+2) \Gamma (q+1)} \\
&=& \nonumber e^{-2\pi y j} \left( \frac{2 \sqrt{2} \pi  \sqrt{\kappa_+}}{\sqrt{y}} + \frac{\sqrt{2} \pi ^2 \kappa_- \sqrt{\kappa_+}}{y^{3/2}}+ \frac{\sqrt{2} \pi ^2 \kappa_+^{3/2} (8 \pi  j y+1)}{3 y^{3/2}} + \cdots \right).
\end{eqnarray}
$a_j(y)$ is given in terms of the spectral density $\rho(j,e)$ by $a_j(y) = \int de \rho(j,e)e^{-2\pi y e}$. Hence, having  $a_j(y)$ we need to take the inverse Laplace transform.  We can invert term by term in the above expansion by using 
\begin{eqnarray}
\label{ILTy}
{\cal L}^{-1}\{ \frac{e^{-2\pi y j} }{y^m} \}= 2\pi \frac{(2\pi)^{m-1}(e-j)^{m-1}}{\Gamma(m)} \Theta(e-j),
\end{eqnarray}
which leads to
\begin{eqnarray}
\label{expansion}
\rho^{s=1}_{(E,J)}(e,j)&=& \left( \frac{4 \pi  \sqrt{\kappa_+}}{\sqrt{e-j}} + \frac{8 \pi ^3 \sqrt{\kappa_+} (e (3 \kappa_-+\kappa_+)+j (\kappa_+-3\kappa_-))}{3 \sqrt{e-j}} + \cdots \right) \Theta(e-j),
\end{eqnarray}
where we denote $\rho^{s=1}_{(E,J)}(e,j)$ the spectral density due to the operator $(E,J)$. Notably, the final result can be given in a closed form
\begin{eqnarray}
\label{final}
\rho^{s=1}_{(E,J)}(e,j)&=& \frac{\sinh \left(\sqrt{2} \pi  \zeta  \left(\sqrt{-E-J}+\sqrt{J-E}\right)-\frac{\sqrt{2} \pi  j \left(\sqrt{J-E}-\sqrt{-E-J}\right)}{\zeta }\right)}{\sqrt{e^2-j^2}}\Theta(e-j) \\
& & + \frac{\sinh \left(\frac{\sqrt{2} \pi  j \left(\sqrt{-E-J}+\sqrt{J-E}\right)}{\zeta }-\sqrt{2} \pi  \zeta  \left(\sqrt{J-E}-\sqrt{-E-J}\right)\right)}{\sqrt{e^2-j^2}}\Theta(e-j), \nonumber
\end{eqnarray}
where we have introduced the combination
\begin{equation}
\zeta =\sqrt{e+\sqrt{e^2-j^2}}.
\end{equation}
Let us make the following remarks. In our derivation we have used $j>0$. This is an arbitrary choice: the spectrum is actually invariant under $j \to -j$ together with $J \to -J$, so that we can recover the answer for $j<0$ from the answer above. On the other hand, for $j=0$ some of the intermediate integrals are not convergent, so the result is not to be trusted. Second, note that in the censored region either $E+J<0$ or $E-J<0$. In this case the density will generically grow exponentially for large quantum numbers.  Furthermore, note that the combinations $E \mp J$ control the behaviour around the essential singularities at $x = \pm i y$. An operator in the censored region is the equivalent of a polar term in the holomorphic case. As for the case of holomorphic forms, other essential singularities can be treated in exactly the same way. Our final expression for the spectral density for finite spin $j$ is then
\begin{eqnarray}
\label{finalRademacher}
\rho_{(E,J)}(e,j)&=& \sum_{s=1,2,\cdots} \sum_{\substack{0<r \leq s,\\(r,s)=1}} e^{2\pi i ( - \frac{r}{s} j + \frac{a[r,s]}{s} J)} \frac{\rho^{s=1}_{(\frac{E}{s^2},\frac{J}{s^2})}(e,j)}{s}\\
&=& \sum_{s=1,2,\cdots} \frac{K(j,J,s)}{s} \rho^{s=1}_{(\frac{E}{s^2},\frac{J}{s^2})}(e,j),
\end{eqnarray}
where $K(j,J,s)$ is the Kloosterman sum, introduced in section \ref{examples}. We would like to claim the sum over $s$ is actually convergent. As can be seen from the explicit answer
\begin{equation}
\rho^{s=1}_{(\frac{E}{s^2},\frac{J}{s^2})}(e,j) \sim \frac{1}{s}.
\end{equation}
The convergence rate exactly agrees with that of the Rademacher expansion for the $j-$invariant function, so that the expression for the density converges for all spin $|j|>0$. Although we have considered a single operator in the censored region, we could consider any finite number, or even a density of them, provided this density does not modify the behaviour at the essential singularities.\footnote{A counterexample of this behaviour is an accumulation point in the twist, where the number of operators grows exponentially with the spin.} In the next section we will study an example of this.  

We conclude this section by noting that our formula is consistent with the known asymptotic behaviour for large quantum numbers $(e,j)$. We can consider different regimes. In the Cardy regime $e \gg j$ we obtain
\begin{equation}
\label{asympte}
\rho^{s=1}_{(E,J)}(e,j) \sim \frac{e^{2 \pi  \sqrt{e} \left(\sqrt{-E-J}+\sqrt{J-E}\right)}}{2 e},
\end{equation}
which agrees with Cardy asymptotics. In the large spin limit $e,j \gg 1$ with $e-j=2\bar h -2\hat c$ fixed we obtain
\begin{equation}
\rho^{s=1}_{(E,J)}(e,j) \sim \frac{e^{2 \sqrt{2} \sqrt{-E-J} \sqrt{j}}}{2 \sqrt{(\bar h-\hat c)(j+ \bar h - \hat c)}}.
\end{equation}
This precisely agrees with the asymptotic behaviour at large spin found in \cite{Benjamin:2019stq}. Note however that this exponential behaviour receives power law suppressed corrections, but with the same exponential factor.  

\section{Ambiguities, comparison to MWK and negative norms}
\label{PvsR}
\subsection{Ambiguities}
We have found a spectral density consistent with the full modular invariance and the presence of an arbitrary operator in the censored region. This answer is not unique, and we could have added to the resulting partition function $Z^p(q,\bar q)$ any Modular invariant function $Z_{amb}(q,\bar q)$ with the following decomposition
\begin{equation}
Z_{amb}(q,\bar q)  = y^{1/2} \sum_{k,\bar k \geq 0} q^{k} \bar q^{\bar k}.
\end{equation}
In the holomorphic case this problem does not arise: One can prove that a bounded modular form of weight zero is necessarily a constant. This makes it possible for the Rademacher sums to reconstruct an entire modular form, of appropriate weight, from its principal part unambiguously. The case of real modular forms is much more complicated, and the space of real forms with specific weights is much harder to characterise, see \cite{Brown:2017qwo}. An example of such an ambiguity is given by the Poincare series starting from an operator in the uncensored region: This is clearly modular invariant and produces an spectrum only in the uncensored region. While we don't have a mathematical proof of such a statement, let us assume this provides a basis for the possible ambiguities. This has a very interesting implication: recall that a seed operator with quantum numbers $(E,J)$ produces an asymptotic density in the Cardy regime of the form (\ref{asympte})
\begin{equation}
\rho^{s=1}_{(E,J)}(e,j) \sim e^{2 \pi  \sqrt{e} \left(\sqrt{-E-J}+\sqrt{J-E}\right)}.
\end{equation}
Since being in the uncensored region implies $E+|J|>0$, the ambiguities above lead to an oscillatory behaviour in the Cardy regime, and hence we conclude the ambiguity does not spoil Cardy behaviour. The same is true for the asymptotic density in the large spin limit. Note that this could be violated  if we have a continuous density of `ambiguous seed operators' which grows exponentially for large $e,j$. On the other hand, this would produce, by modular invariance, some operators in the uncensored region. Hence, we assume the density of ambiguous seed operators grows at most polynomially. This would not spoil the asymptotic Cardy behaviour. Another important constraint arises from {\it positivity} of the spectrum: in a unitary CFT the density of states should be positive. In our framework positivity needs to be imposed `by hand'. In particular we are free to add the ambiguous terms above (which are oscillatory) provided we don't spoil this positivity, or conversely, may need to add them, to make the density positive. It would be interesting to understand precisely how this constraints the ambiguities.   

\subsection{Comparison to MWK}

Maloney, Witten and Keller (MWK) constructed a family of modular invariant partition functions consistent with the presence of a seed operator $(E,J)$. In the following, we would like to compare their result to ours: we will show that the partition functions that follow from the Rademacher expansions are an interesting modification of the ones constructed by  MWK. The densities found by MWK have been summarised in the appendix. Let us analyse first the case of a scalar seed. As reviewed in the appendix the MWK  density can be written as
\begin{equation}
\rho_{_{P}}(e,j) = \sum_{s=1}^\infty \frac{1}{s} K(j,0;s)  \sum_{m=1,2,3,\cdots} \frac{|j|^{m-1}}{s^{2m}} c_m (-E)^m T_m(e/|j|),
\end{equation}
where $T_q(t)=\cosh( q \cosh^{-1} t)$ denote the Chebyshev polynomials of the first kind.  The density is denoted by $\rho_{_{P}}(e,j)$ to recall that it arises from a Poincare construction. We find the following remarkable result: for a scalar seed the density found in this paper can be written in a very similar fashion
\begin{equation}
\rho_{_{R}}(j,\Delta) = \sum_{s=1}^\infty \frac{1}{s} K(j,0;s)  \sum_{m=\frac{1}{2},\frac{3}{2}, \frac{5}{2}, \cdots} \frac{|j|^{m-1}}{s^{2m}} c_m (-E)^m T_m(e/|j|),
\end{equation}
where now the sum over $m$ runs over half-integers. To determine that this is also the case for a non-scalar seed, with $J \neq 0$, is non-trivial, as one needs to analytically continue the expression for $D_t^m T_m(t)$ to non-integer $m$. This can be done in an expansion around $t=1$:
\begin{equation}
D_t^m T_m(t) = \frac{(-E-J)^m}{\sqrt{2} \sqrt{t-1}}+ \frac{\left((2 m-1) (-E-J)^m (2 E m+E-2 J m+J)\right)}{4 \sqrt{2} (E+J)}\sqrt{t-1} + \cdots,
\end{equation}
where this series can be continued to very high order in $t-1$. Given this expression, the sum over $m$ can be performed order by order in $t-1$. When summing over integer $m$ this agrees with the expression given in (\ref{MWKdensity}), while if we sum over half-integer $m$ we obtain precisely (\ref{final}).

Let us discuss the physical difference between both constructions. The MWK partition function is modular invariant by construction, since it corresponds to the sum over images of a seed operator. On the other hand, given a Rademacher expansion it is not always clear that it lands on a modular invariant form.\footnote{Although convergent, the expansion may land on a Mock modular form.}  To study this question let us focus in a scalar seed. In this case the MWK partition function can be written as a sum of well known real analytic modular forms, denoted the Eisenstein series
\begin{equation}
E(\tau,s) = \frac{1}{2} \sum_{(m,n)=1} \frac{y^s}{|m \tau+n|^{2s}}.
\end{equation}
These have the following Fourier decomposition
\begin{equation}
E(\tau,s) = y^s + \frac{\hat \zeta(2s-1)}{\hat \zeta(2s)} y^{1-s} + \frac{4}{\hat \zeta(2s)} \sum_{j=1} \cos(2\pi j x)j^{s-1/2} \sigma_{1-2s}(j) y^{1/2} K_{s-1/2}(2\pi j y),
\end{equation}
where $\sigma_{1-2s}(j)$ is the divisor function and 
\begin{equation}
\hat \zeta(s) = \pi^{-s/2} \Gamma \left( \frac{s}{2}\right) \zeta(s).
\end{equation}
The total partition function $Z_{_{MWK}}^p(\tau)$ is then given by 
\begin{equation}
Z_{_{MWK}}^p(\tau) = \sum_{m=1} \frac{(2 \pi )^m (-E)^m}{\Gamma (m+1)} E(\tau,m+\frac{1}{2}).
\end{equation}
The seed operator is recovered by keeping the first Fourier coefficient in the Eisenstein series  $E(\tau,s) = y^s + \cdots$, resuming this we obtain
\begin{equation}
Z_{_{MWK}}^p(\tau) = \sqrt{y} e^{-2 \pi E y}  + \text{images},
\end{equation}
which exactly corresponds to a scalar seed operator with $E= \Delta_0-2\hat c$. The partition function obtained by the Rademacher procedure is also a linear combination of Eisenstein series, and hence (formally) modular invariant, where now we sum over half-integer $m$.\footnote{Of course this includes the Eisenstein series $E(\tau,1)$ which is not well defined. We will discuss this below.} The corresponding `seed' operator would be
\begin{equation}
Z_{_R}^p(\tau) =\sqrt{y} e^{-2 \pi  E y} \text{erf}\left(\sqrt{2 \pi } \sqrt{-E y}\right) + \text{images}.
\end{equation}
This corresponds to the usual seed operator, in the censored region, together with a density of extra scalar operators in the uncensored region (with $E>0$) such that
\begin{equation}
\int_0^\infty e^{-2 E' \pi y} \rho_{extra}(E') dE' = e^{-2 \pi  E y} \left( \text{erf}\left(\sqrt{2 \pi } \sqrt{-E y}\right) -1 \right).
\end{equation}
We find 
\begin{equation}
\label{extraops}
\rho_{extra}(E')  = \frac{\sqrt{-E}}{\pi (E-E')\sqrt{E'}} \Theta(E').
\end{equation}
This is an example of the ambiguities mentioned above. Hence, both constructions are equivalent up to this extra contribution. We expect something similar to happen for $J \neq 0$. It is tantalising to conjecture that this is the sort of ambiguities expected for generic non-rational CFTs. 

There is an alternative way to understand (\ref{extraops}). We have seen that (\ref{final}) gives the density of states that follows from the Rademacher construction, while (\ref{MWKdensity}) gives the density that follows from the Poincare series construction. For simplicity lets work in the regime of small $\epsilon=e-j$. For a scalar seed of dimension $E$  
\begin{equation}
\rho_{_{R}}(j) \sim \frac{\sqrt{2} \sinh \left(2 \sqrt{2} \pi  \sqrt{-E j}\right)}{\sqrt{j} \sqrt{\epsilon }},~~~\rho_{_{P}}(j) \sim \frac{2 \sqrt{2} \sinh ^2\left(\sqrt{2} \pi  \sqrt{-E j}\right)}{\sqrt{j} \sqrt{\epsilon }},
\end{equation}
where higher orders in $\epsilon$ are ignored and only the leading $s=1$ contribution is considered. Let us now imagine we have a density of seeds given by (\ref{extraops}). The total contribution of their Poincare series to the density is then
\begin{equation}
\int_0^\infty \frac{2 \sqrt{2} \sinh ^2\left(\sqrt{2} \pi  \sqrt{-E' j}\right)}{\sqrt{j} \sqrt{\epsilon }}\frac{\sqrt{-E}}{\pi (E-E')\sqrt{E'}} dE' =\frac{\sqrt{2}(1-e^{-2 \sqrt{2} \pi \sqrt{-E j }})}{\sqrt{j \epsilon }},
\end{equation}
which exactly accounts for the difference between $\rho_{_{R}}(j)$ and $\rho_{_{P}}(j)$. 

\newpage

\noindent{\bf Extension to $j=0$}

\bigskip

\noindent The Rademacher expansion does not converge for $j=0$. A natural question is whether modular invariance may be used to define an extension.  We have seen that in the case of a scalar light operator the density of states with $j>0$ arises from a partition function which is formally a sum over real Eisenstein series 
\begin{equation}
Z_{_{R}}^p(\tau) = \sum_{m=\frac{1}{2},\frac{3}{2},\cdots} \frac{(2 \pi )^m (-E)^m}{\Gamma (m+1)} E(\tau,m+\frac{1}{2}).
\end{equation}
In the $s-$plane the real Eisenstein series $E(\tau,s)$ is a meromorphic function with a pole at $s=1$
\begin{equation}
E(\tau,s) = \frac{\pi}{s-1} + \cdots.
\end{equation}
This pole gives a divergent contribution that affects the density of states only for $j=0$. This is the counterpart of the situation in the MWK construction, where a regularisation procedure is necessary. It would be interesting to understand whether the Rademacher construction should naturally land on a specific density for $j=0$, as for the holomorphic case, see example 1 in section \ref{examples}.   

\subsection{Negative densities}
\label{negativedensity}
Having the densities (\ref{final}) and  (\ref{MWKdensity}) that follow from the Rademacher and Poincare constructions we can study several quantitative questions. Let us study the issue of negative norm states in the CFT dual of pure gravity on $AdS_3$, first observed in \cite{Benjamin:2019stq}. In its simplest version the CFT is given by a unitary 2d $CFT$ at large central charge whose operators consist of the vacuum plus a tower of operators with $h,\bar h \geq \hat c$, corresponding to the BTZ black holes.
The partition function of the putative CFT must contain the identity operator, no other Virasoro primaries outside the region $h,\bar h \geq \hat c$, and must be consistent with modular invariance. A candidate for such a partition function is the Poincare construction of MWK \cite{Maloney:2007ud,Keller:2014xba}, while another equally good construction is given by the Rademacher construction considered in this paper. A problem with both partition functions is that they don't posses a discrete spectrum. Another problem is the existence of negative norm states, observed in the regime of large spin by \cite{Benjamin:2019stq}. In the following we will show that such states are also present at finite spin (as small as one), and discuss scenarios to cure this negativity.  

Let us consider the contribution to the density $\rho(e,j)$ from the vacuum operator. Due to the form of the vacuum character, this is given by the sum of four terms, each of the form (\ref{final}) or (\ref{MWKdensity}):
\begin{equation}
\rho_{vac}(e,j) =  \rho_{(-2\hat c,0)}(e,j)- \rho_{(1-2\hat c,1)}(e,j)-\rho_{(1-2\hat c,-1)}(e,j)+ \rho_{(2-2\hat c,0)}(e,j).
\end{equation}
We now consider the density $\rho_{vac}(e,j)$ in the region of positive spin $j$ and small $e-j =\epsilon$. When plugging this expression into (\ref{finalRademacher}) we find the leading contribution, with $s=1$, vanishes as $\sqrt{\epsilon}$:
\begin{equation}
\rho_{vac}^{s=1}(e,j) \sim \frac{4 \sqrt{2} \pi ^2\left(\sinh \left(4 \pi  \sqrt{\hat c j}\right)-\sinh \left(4 \pi  \sqrt{(\hat c-1) j}\right)\right)}{\sqrt{j}}\sqrt{\epsilon },
\end{equation}
while the next contribution arises from $s=2$ and equals
\begin{equation}
\rho^{s=2}_{vac}(e,j) =  \frac{\sqrt{2} e^{-i \pi  j} \left(\sinh \left(2 \pi  \sqrt{(\hat c-1) j}\right)+\sinh \left(2 \pi  \sqrt{\hat c j}\right)\right)}{\sqrt{j} \sqrt{\epsilon }} +{\cal O}(\sqrt{\epsilon }),
\end{equation}
with similar behaviour for $s=3,4,\cdots$. Note that for any odd spin $j$ we can always choose $\epsilon$ sufficiently small such that the second contribution overcomes the first, and the total density is negative. Exactly the same conclusions are reached if we analyse the Poincare density (\ref{MWKdensity}). This agrees with the asymptotic analysis  in \cite{Benjamin:2019stq} but the formulas presented here are valid for finite values of the spin, modulo the ambiguities mentioned above. A natural question is whether these ambiguities can cure the negativity of the density. According to our discussion at the beginning of this section, this could happen in a regime where the negative density does not grow exponentially. However, this negative density grows exponentially either at large spin, or large central charge, and this cannot be cured by the ambiguities. Note that in our discussion we have assumed there is no accumulation points in the twist or the accumulation is mild enough. We can put this in a different way:  Since we got a continuous density, we need to check that the `amount' of operators between $e=j$ and $e=j+\epsilon$ with negative norm is exponentially large, otherwise this can be cured by the ambiguities. Let us estimate the regime at which the norm becomes negative. The over-crossing happens when 
\begin{equation}
\epsilon \sim e^{-2 \pi \sqrt{\hat c j}}.
\end{equation}
In the regime in which the ambiguities are not relevant $\hat c j$ is large, so that $\epsilon$ is exponentially small. The amount of operators in the appropriate regime is then
\begin{equation}
\int_{j}^{j+\epsilon_0} \rho_{\epsilon}d\epsilon \sim \int_{j}^{j+\epsilon_0} \frac{e^{2 \pi \sqrt{\hat c j}}}{\sqrt{j \epsilon}} d\epsilon \sim e^{\pi \sqrt{\hat c j}},
\end{equation}
which is an exponentially large number of operators. 

Let us study a few scenarios to cure this negativity. First, as in \cite{Benjamin:2019stq}, we can add extra isolated operators in the censored region. For this consider the density $\rho^{s=1}_{(E,J)}(e,j)$ in the region under consideration. From (\ref{final}) we obtain
 \begin{equation}
 \rho^{s=1}_{(E,J)}(e,j)  = \frac{\sqrt{2} \sinh\left(2 \sqrt{2} \pi  \sqrt{-j (E+J)}\right)}{\sqrt{j} \sqrt{\epsilon }} + {\cal O}(\sqrt{\epsilon}).
 \end{equation}
Its asymptotic form for large $j$ agrees with that given in \cite{Benjamin:2019stq}, but this expansion is valid even for finite spin, again modulo the ambiguities mentioned above. Denoting $E+J = 2\tau-2\hat c$, with $\tau$ the twist of the new operator, we see that in order to cancel the above negativity we need $\tau=\frac{3}{4} \hat c$, or smaller. Let's assume $\tau=\frac{3}{4} \hat c$. Is this enough? With our formulas we may ask if this cures the negativity down to finite odd spin. This turns out not to be the case: we need to add another operator with  $\tau_2 =\frac{3}{4} \hat c + \frac{1}{4}$, and then another one with $\tau_3= \frac{8}{9} \hat c$. After this, the amount of operators with negative norm is of order one, and this could in principle be cured by ambiguities. If we add another operator of twist $\tau_4= \frac{15}{16} \hat c$, then the amount of operators with negative norm is exponentially small.\footnote{The fact that states with twists $\tau_n$ can be used to cure the negativity in the density of states was independently observed by Benjamin, Collier and Maloney, who will discuss the physical interpretation of this fact in \cite{BCM}.} 

Let us consider another scenario where we have an accumulation in the twist. Namely, we consider a tower of operators with fixed twist $\tau_0 =min(h_0,\bar h_0)$ and spin $J=h_0-\bar h_0$ ranging from minus to plus infinity. Recall $E=h_0+\bar h_0-2\hat c$. For positive spin $\tau_0 =\bar h_0$ and $E=2 \tau_0+J-2\hat c$. For negative spin $\tau_0 =h_0$ and $E=2 \tau_0-J-2\hat c$, so that $E=E_0+|J|$ with $E_0 = 2 \tau_0-2\hat c$. In order to run our procedure for this case we need to consider the appropriate sum on the r.h.s. of (\ref{Rademachersingle}). From (\ref{ILTy}) it follows that in the regime of small $\epsilon = e-j$ the main contribution to the density of states arises from the region of large $y$. In this limit we obtain
\begin{equation}
\sum_{J=-\infty}^\infty \frac{2 e^{j(\frac{1}{w}-2\pi y)}}{w \sqrt{-1+4 \pi w y}} e^{\frac{4 \pi ^2 w (1-2 \pi  w y)}{4 \pi  w y-1} J} e^{-\frac{8 \pi ^3 w^2 y}{4 \pi  w y-1}(E_0+|J|)} \sim e^{j(\frac{1}{w}-2\pi y)} e^{-2 \pi^2 w E_0} \frac{\sqrt{y}}{w^{3/2}}.
\end{equation}
This is to be contrasted for the behaviour for a single operator which behaves as $y^{-1/2}$ instead. From this we conclude that in the small $\epsilon$ limit the accumulation in the twist under consideration produces a density of the form 
 \begin{equation}
 \rho^{s=1}_{acc}(\Delta,j)  \sim \frac{ \sinh\left(2 \sqrt{2} \pi  \sqrt{-j E_0 }\right)}{\sqrt{j} \epsilon^{3/2}}.
 \end{equation}
Hence, the presence of a tower of operators results in an enhancement in the behaviour of the density as $e-j \to 0$. We can consider a generalisation of this toy model, where we have the same tower of operators, but with degeneracy $|J|^d$ for spin $J$. For $d=1,2,\cdots$ this degeneracy gives an even greater enhancement as $\epsilon \to 0$, which is easily computable:
 \begin{equation}
 \rho^{s=1}_{acc}(\Delta,j)  \sim \frac{ \sinh\left(2 \sqrt{2} \pi  \sqrt{-j E_0 }\right)}{\sqrt{j} \epsilon^{3/2+d}}.
 \end{equation}
Of course, these toy models, even for $d=0$, give a non-integrable density of states, so they do not provide a satisfactory solution to the negativity problem, but they show that our framework can be used even in situations with accumulation points in the twist. 

\section{Conclusions}

In the present paper we have considered the spectrum of operators in a unitary 2d CFT with Virasoro symmetry. Modular invariance of the partition function imposes strong constraints, and we have constructed a density $\rho_R(e,j)$ consistent with full modular invariance and the presence of  a given spectrum in the censored region. The density is given by a Rademacher expansion, which is convergent for $|j| \neq 0$. For large spin it reproduces the correct asymptotic behaviour while for finite spin it is given by a variant of the Poincare construction developed by Maloney, Witten and Keller. It would be interesting to explore the following directions. 
\begin{itemize}
\item It would be interesting to study further the explicit expressions we have derived for $\rho_R(e,j)$ and $\rho_P(e,j)$, specially in the limit of large central charge and a sparse light spectrum. In particular one should be able to reproduce the results of \cite{Hartman:2014oaa}, but also improve them, since we took into account the full modular invariance. 
\item We have seen that the densities  $\rho_R(e,j)$ and $\rho_P(e,j)$ are defined up to ambiguities, given by modular invariant corrections with a spectrum purely in the uncensored region. It would be interesting to show mathematically that such ambiguities are always given by the Poincare series generated from seeds in the uncensored region. This is physically very natural, and it warrantees the asymptotic behaviour is not spoilt. Along the same lines, it would be interesting to restrict such ambiguities. As mentioned above, imposing positivity of the spectrum should give strong restrictions on the ambiguities we can add. 
\item We have studied a scenario to cure the negativity of the `pure gravity' density. This scenario suggests the inclusion of a tower of operators of twist $\tau_1,\tau_2,\cdots$ given in section \ref{negativedensity}. It would be interesting to study the resulting partition function, and in particular the discreetness of the resulting density. 
\item We have studied toy models with accumulation in the twist. Since they result in non-integrable densities, they don't provide a solution to the negativity problem, but they show that accumulation points in the twist can be studied in our framework. It would be interesting to study more general models.  
\item In \cite{Maxfield:2019hdt} quantum corrections to the extremality bound of the BTZ black hole were studied. It would be interesting to see if these quantum corrections modify the statements about negativity of the spectrum.  
\item It would be interesting to consider scenarios where the spectrum is discrete. We expect the spectrum to become `dense' in the large $c$ (or large spin) limit, with $e^{-\alpha(j) \sqrt{c}}$ the typical separation between operators, but discreteness should reappear at finite $c$ and finite $j$. In this case the integrated density may be a better observable, and Tauberian theory , studied for instance in see \cite{Mukhametzhanov:2019pzy}, seems to be the appropriate framework. 
\item It would be interesting to combine our results with the techniques of  \cite{Hellerman:2009bu,Friedan:2013cba,Collier:2016cls,Bae:2017kcl} in order to explore bounds in the spectrum of operators in the censored region. Note that we have made use of the full modular invariance of the partition function, while these works have obtained impressive results using only a subset, together with positivity. 
\item Along similar lines  it would be interesting to make contact with the sphere packing problem shown to be related to the modular bootstrap program in \cite{Hartman:2019pcd}.
\item It would be interesting to combine our techniques/results with the study of four-point correlators in the context of the modular bootstrap, see \cite{Kusuki:2018wpa,Collier:2018exn}. 
\end{itemize}

\section*{Acknowledgements} 
We would like to thank Miranda Cheng and Alex Maloney for helpful discussions. This project has received funding from the European Research Council (ERC) under the European Union's Horizon 2020 research and innovation programme (grant agreement No 787185).

\appendix

\section{MWK density of states}
Maloney, Witten and Keller (MWK) constructed a family of modular invariant partition functions  \cite{Maloney:2007ud,Keller:2014xba} labelled by a seed operator with quantum numbers $(E,J)$, with $E=h_0+\bar h_0-2\hat c$. The partition function is given by an appropriately regularised Poincare series obtained from the seed by summing over all its $PSL(2,\mathbb{Z})$ images. For each integer spin $j$ the spectral density $\rho_{_{P}}(e,j)$, where $P$ stands for Poincare, is a continuous function of $e=\Delta-2\hat c$. We say an operator belongs to the censored region if either $e+j<0$ or $e-j<0$. The uncensored region is the complement of this, namely $e -|j|>0$. It turns out that if the seed is in the uncensored region  then the spectrum has support only in the uncensored region, while if the seed is in the censored region, that is, either $E-J<0$ or $E+J <0$, then the seed is the only operator in the censored region. The spectral density can be written as a sum of elementary contributions 
\begin{equation}
\rho_{_P}(e,j) = \sum^{\infty}_{m=0} \rho_m(e,j).
\end{equation}
The expression for the individual contributions $\rho_m(e,j)$ depends on whether the seed is scalar or not, and whether $j=0$ or not. In order to compare the MWK spectral density to ours we will be interested in the case $j \neq 0$. Let us treat first the case $J=0$ and then the general case. 

\bigskip

\noindent{{\bf Scalar seed}}

\bigskip

\noindent In this case and for $j \neq 0$
\begin{equation} 
\label{scalarseedjneq0}
\rho_{0}(e,j) = 0,~~~\rho_{m}(e,j)  = \frac{\sigma_{2m(j)}}{|j|^{m+1}\zeta_{2m+1}} c_m (-E)^m T_m(e/|j|) \Theta(e-|j|),~~~m=1,2,\cdots,
\end{equation}
where $\sigma_{2m}(j)$ is the divisor function, $T_q(t)$ denote the Chebyshev polynomials of the first kind
\begin{equation} 
T_q(t) = \cosh(q \cosh^{-1} t),
\end{equation}
and we have introduced
\begin{equation} 
c_m = \frac{2^{2m+1}\pi^{2m}}{\Gamma(2m+1)}.
\end{equation}
We would like to find a close form expression for the density $\rho_{_{P}}(e,j)$ in this case, so that we can make a comparison to our results. For this rewrite the divisor function $\sigma_{2m(j)}$ in terms of the Kloosterman sum, see for instance \cite{Keller:2014xba} appendix B:
\begin{equation}
\sigma_{2m}(j) = j^{2m} \zeta_{2m+1} \sum_{s=1}^\infty s^{-2m-1} K(j,0;s),
\end{equation}
then
\begin{equation}
\rho_{_{P}}(e,j) = \sum_{s=1}^\infty \frac{1}{s} K(j,0;s)  \sum_{m=1} \frac{|j|^{m-1}}{s^{2m}} c_m (-E)^m T_m(e/|j|)  = \sum_{s=1}\frac{1}{s} K(j,0;s) \rho^{(s)}_{_{P}}(e,j),
\end{equation}
where we have suppressed $\Theta(e-|j|)$ for convenience. For each fixed $s$ the sum over $m$ can be performed. We obtain

\begin{equation}
 \rho^{(s)}_{_{P}}(e,j)  = \frac{\cosh \left(\frac{2 \sqrt{2} \pi  j \sqrt{\frac{-E}{\sqrt{e^2-j^2}+e}}}{s}\right)+\cosh \left(\frac{2 \sqrt{2} \pi  \sqrt{-E \left(\sqrt{e^2-j^2}+e\right)}}{s}\right)-2}{\sqrt{e^2-j^2}} \Theta(e-j).
\end{equation}

\bigskip

\noindent{{\bf Non-scalar seed}}

\bigskip

\noindent In this case and for $j \neq 0$
\begin{equation} 
\label{nonscalarseedjneq0}
\rho_{m}(e,j)  = Z_{j,J}(m+1/2)c_m |j|^{m-1} D_t^m T_m(e/|j|) \Theta(e-|j|),~~~m=1,2,\cdots,
\end{equation}
where $Z_{j,J}(m+1/2)$ is the Kloosterman zeta function
\begin{equation}
Z_{j,J}(m+1/2) = \sum_{s=1}^\infty s^{-2(m+1/2)} K(j,J;s),
\end{equation}
and $D_t$ is given by a derivative operator acting on the Chebyshev polynomials. More precisely
\begin{equation} 
D_t T_m(t)  = (-E +J (\partial_t + \frac{m}{t}))T_m(t).
\end{equation}
The resulting density can be be computed as a power expansion in $E$ and $J$. At each order, only a finite number of $\rho_{m}(e,j)$ contributes. The expressions became very cumbersome very soon, but computing a large number of terms we were able to guess their general form. They resummed in the following expression for the density
\begin{eqnarray}
\label{MWKdensity}
& & \rho^{s=1}_{_P}(\Delta,j)=\frac{\cosh \left(\sqrt{2} \pi  \zeta  \left(\sqrt{-E-J}+\sqrt{J-E}\right)-\frac{\sqrt{2} \pi  j \left(\sqrt{J-E}-\sqrt{-E-J}\right)}{\zeta }\right)-1}{\sqrt{e^2-j^2}}\Theta(e-j) \nonumber \\
& &+ \frac{\cosh \left(\frac{\sqrt{2} \pi  j \left(\sqrt{-E-J}+\sqrt{J-E}\right)}{\zeta }-\sqrt{2} \pi  \zeta  \left(\sqrt{J-E}-\sqrt{-E-J}\right)\right)-1}{\sqrt{e^2-j^2}}\Theta(e-j),
\end{eqnarray}
where we have introduced the combination
\begin{equation}
\zeta =\sqrt{e+\sqrt{e^2-j^2}}.
\end{equation}
In finding this result we have excluded the contribution from $m=0$, otherwise one would not have the $``1"$ in the expression above. On the other hand, excluding this contribution makes the limit $J \to 0$ smooth.

\end{document}